%% file: main.tex
\documentclass[11pt]{article}
\usepackage[utf8]{inputenc}
\usepackage[margin=1in]{geometry}
\usepackage{comment}
\usepackage{natbib}
\usepackage{hyperref}
\hypersetup{linkcolor=blue, citecolor=blue}
\usepackage{multicol}
\usepackage{graphicx}
\usepackage{subfig}
\usepackage{wrapfig}
\usepackage[font=small,labelfont=bf]{caption}
\usepackage{amssymb}
\usepackage{color, soul}
\usepackage[superscript, biblabel]{cite}

\newcommand{\apjl}{\textit{Astrophysical Journal Letters}}
\newcommand{\apj}{\textit{The Astrophysical Journal}}
\newcommand{\aj}{\textit{The Astronomical Journal}}
\newcommand{\nat}{\textit{Nature}}
\newcommand{\mnras}{Monthly Notices of the Royal Astronomical Society}

\newcommand{\apjs}{The Astrophysical Journal Supplement}
\newcommand{\aap}{\textit{Astronomy \& Astrophysics}}
\newcommand{\icarus}{\textit{Icarus}}
\newcommand{\pasp}{\textit{Publications of the Astronomical Society of the Pacific}}
\newcommand{\pasj}{\textit{Publications of the Astronomical Society of Japan}}
\newcommand{\aaps}{\textit{Astronomy \& Astrophysics, Supplement}}
\newcommand{\psj}{\textit{The Planetary Science Journal}}
\newcommand{\araa}{\textit{Annual Review of Astronomy and Astrophysics}}
\newcommand{\na}{\textit{New Astronomy}}

\begin{document}
\sloppy

\begin{center}
\begin{LARGE}

Detecting and Characterizing Exomoons and Exorings\\
\end{LARGE}

\vspace{0.5em}

from \textit{The Handbook of Exoplanets}, 2nd edition

\vspace{1.0em}
Alex Teachey\\
Academia Sinica Institute of Astronomy \& Astrophysics\\
11F of AS/NTU Building, No. 1, Sec. 4, Roosevelt Rd, Taipei 10617, Taiwan (R.O.C.)

\end{center}
\vspace{1.0em}

\begin{center}
    \begin{large}
    ABSTRACT
    \end{large}
\end{center}
The circumplanetary environments in our Solar System host a stunning array of moon and ring systems. Study of these environs has yielded valuable insights into planetary system formation and evolution, and there is every reason to believe that we will have much to learn from the moons and rings that are likely to exist in exoplanetary systems as well. This has motivated a small but growing number of researchers to investigate questions related to the formation, stability, long-term viability, composition, and observability of these exomoons and exorings. Still, due to a range of significant observational challenges, we remain at a relatively early stage of this work. As a result, we continue to face a number of difficult, unanswered questions, but this also means there are myriad opportunities for fundamental contributions to the field. In this review we will examine a variety of important issues for the astronomical community to consider, with an aim of providing a comprehensive understanding of ongoing efforts to identify and characterize exomoons and exorings, while also increasing interest and engagement. We begin with an overview of what we expect from systems hosting moons and/or rings in terms of their architectures, habitability, and observational signatures. We then highlight the contributions from a variety of works that have been aimed at detecting and characterizing them. We conclude by examining the outlook for finding these objects and discussing a number of ongoing challenges that we will want to overcome in the years ahead.

\input{introduction.tex}

\input{motivations.tex}

\input{predictions.tex}

\input{detection.tex}

\input{characterization.tex}

\input{challenges.tex}

\input{conclusion.tex}

\input{xref.tex}

\input{bibliography.tex}

\end{document}

%% file: introduction.tex
\label{sec:introduction}

\section{Introduction}
The catalogue of known exoplanets has seen enormous growth over the last two decades, yielding important insights into the formation and evolution of planetary systems. The identification of exotic system architectures, breakthroughs in atmospheric characterization, and the ability to contextualize our Solar System among a large population of other systems, have all revolutionized our understanding of these processes. Still, there remain areas in which our knowledge of exoplanetary systems lags far behind that of our own Solar System.
This is to be expected, as some tools routinely used for Solar System research are simply not available for more distant systems; we are sadly unable to visit the exoplanets to observe them close-up. Yet there remains a great deal more we can learn about these systems even as we are restricted to long range photometric and spectroscopic studies. In particular, the circumplanetary environments -- which in our Solar System display a stunning array of moons and ring systems -- remain largely uncharacterized in exoplanetary systems.

\vspace{0.5em}

The last several years have seen remarkable advances in the related fields of %\hl{exomoons} 
exomoons and %\hl{exorings}
exorings, and it is an exciting time to be engaged with this work. This despite the fact that we still lack a single confirmed exomoon detection (though we have some intriguing candidates), and exorings are similarly elusive. Outside observers might naturally wonder, what is taking so long? Have we nothing to show for this effort, now in its second decade? Well of course we do -- even as the \textit{search} for exomoons and exorings has yet to bear much fruit, we have learned a great deal: we continue to identify new phenomena and develop new tools for the search, we have put important constraints on where and to what extent these worlds will be found, and our observational strategies continue to evolve.

\vspace{0.5em}

It is useful, then, to survey the state of the literature again for this new edition, to see how far we have come, and to synthesize a way forward from what we have built for ourselves. Identifying and collating the various ongoing challenges gives us a better shot at finding a way forward, and bringing new scientists on board. And it is absolutely critical that we get more scientists involved; the more brains chewing on this problem, the faster we can expect to make progress. And the more scientists engaged with exomoons and exorings, the more community buy-in we can expect, which routinely translates into more funding, more telescope time, more prioritizing this objective. Hence, this review.

\vspace{0.5em}

It should be noted that the literature on exomoons and exorings, though comparatively small in number, is already substantial in scope. The $\sim200$ papers treating the subjects of exomoons and exorings are a mere fraction of the works that provide important background on these topics: critical studies of the formation, evolution, dynamics, and composition of moons and rings in our Solar System have a far deeper history. We cannot possibly cover or even acknowledge all of those contributions to our present understanding. A sincere effort has been made to provide the reader with an extensive overview, and plenty of references with which they may begin to explore these topics in greater depth. 

\vspace{0.5em}

We will begin by briefly examining the motivations for undertaking this work, and what we expect of exoplanetary systems hosting moons or rings in terms of their architectures, habitability, and observational signatures. We will then explore in depth recent efforts to identify and characterize these systems, and discuss the process of vetting a potential discovery. We will end by highlighting some of the lingering challenges with which the community will have to contend as we go forward.

%% file: motivations.tex
\label{sec:motivations}

\section{Motivations}
Exoplanetary science is an inherently comparative field. When we study extrasolar planets, we are not \textit{just} learning about those systems, we are also learning more about ourselves, the ways in which our Solar System is ordinary, and the ways in which it is extraordinary. We see moons and rings in abundance within our own Solar System, but is this a typical outcome of planet formation? 

\vspace{0.5em}

Every planetary system, whether it is our own or another, can be seen as only a snapshot in time. But these are dynamic systems, evolving over the course of billions of years. The story of our own Solar System has been pieced together from a range of extant clues -- for example, the presence of rocky planets close to the Sun and gas giants farther out, the location and structure of the asteroid and Kuiper belts, and of course, the presence and sometimes absence of moons and rings. Each planet's collection or dearth of moons and rings tells a story about that planet's own dynamical history, and its evolution in the context of the Solar System. 

\vspace{0.5em}

Having now discovered enough exoplanets to paint a picture (albeit incomplete) of planetary demographics, we have come to understand that in some ways the Solar System may not be typical. But when it comes to moons and rings, there is hardly any point of comparison so far. Might exomoons and exorings be more common, or less common, than their Solar System counterparts? Considering the potential importance of Earth's moon to live on Earth (e.g. \citealt{moon_important}, but see also \citealt{moon_not_important}), are these sorts of moons formed routinely, or rarely? Might there be moons that are astonishingly different from any found in our Solar System? Could they be much larger, could they be ``Earth-like''? And might they help us understand the dynamical history of their own planetary systems? And what about the exorings... are they long-lived, or transient features? Might they be found in much closer proximity to their host star? Could they be rocky instead of icy? Might they be found circling terrestrial, rather than gaseous, worlds? Might they be confusing astronomers into deducing peculiar (unphysical) properties of some of the planets that we've already discovered?

\vspace{0.5em}

Theory can tell us a great deal, but it must always be tested by observation. To find the truth, we must turn to the telescope, and think creatively if we wish to turn photons into answers.

%% file: predictions.tex
\label{sec:predictions}

\section{Predictions and Expectations}

\begin{figure}
    \centering
    \includegraphics[width=\columnwidth]{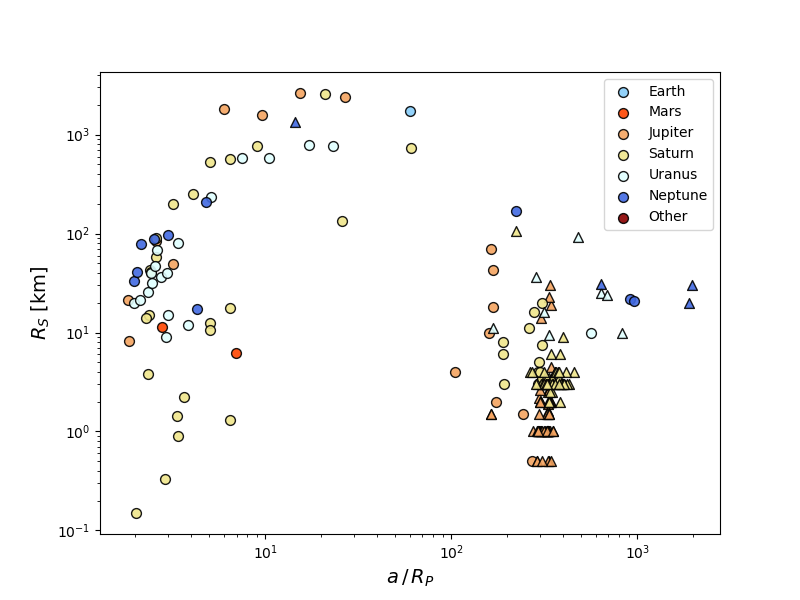}
    \caption{Moons of our Solar System arranged by size and semi-major axis normalized by the planet's radius. Circles indicate prograde moons while triangles are assigned to retrograde moons.}
    \label{fig:solar_system_moons}
\end{figure}

The classic sequence of events for scientific discovery is 1) observe, 2) question, 3) hypothesize, and 4) test. When it comes to moons, we begin with the observation that the Solar System hosts a great diversity of them. In the 400-odd years since Galileo first discovered the four large moons of Jupiter, astronomers have discovered nearly 300 moons orbiting the major planets of our Solar System, mostly small, but a sizeable number which we might describe as worlds -- having achieved hydrostatic equilibrium and with radii of at least several hundred kilometers (Figure \ref{fig:solar_system_moons}). Their abundance, and indeed, the fascinating diversity of geological features we have observed thanks to robotic exploration missions naturally lead us to ask: how did they come to be, and might other planetary systems also host such worlds?

\vspace{0.5em}
We have also discovered that all of the giant planets of our Solar System host rings \citep{Uranus_rings, Io_volcanoes, Neptune_rings, Neptune_ring2}, and even some minor bodies host rings as well \citep{Chariklo_rings, Chiron_rings, Haumea_rings, Quaoar_rings}. But how did these rings come about, and how long do they last? In the case of Saturn's rings, for example, are they as old as the planet itself? Are they the leftovers of a moon-forming circumplanetary disk, a collision of minor bodies, or a tidal disruption event? If so, their structure and composition could tell us a great deal about the primordial environment of Saturn. Or are the rings a short-lived phenomenon, and it is just a happy accident that we live in a brief window of time when they are at their most spectacular \citep[e.g.][]{Saturn_ring_loss}? And if so, what might this say about the likelihood of seeing such ring systems in exoplanetary systems?

\vspace{0.5em}

Armed with a comparatively thorough understanding of moons and rings in our Solar System, their properties and occurrence rates, it is now incumbent on us as a community to look to exoplanetary systems for answers to these same fundamental questions, making hypotheses and testing them. We will start with what we expect from these systems.

\subsection{Formation Pathways and System Architectures}

\subsubsection{Moon formation}
There is now a rich body of literature that seeks to explain the moons we see in our Solar System. Broadly speaking, moons are thought to form in one of three ways. Leaving aside the relative likelihood of these scenarios, they are in no particular order: 1) a giant impact \citep[e.g.][]{Hartmann:1975, Lucey:1995, Canup:2001, Canup:2004, Canup:2012}, 2) formation in a circumplanetary disk \citep[e.g.][]{Lunine:1982, Canup:2006, Ogihara:2012, Heller:2015, Cilibrasi:2018, Szulagyi:2018, Batygin:2020}, and 3) capture of initially unbound objects \citep[e.g.][]{Pollack:1979, Agnor:2006, Nesvorny:2007, Hamers:2018, Hansen:2019}. There may be some gray area between these three pathways; after all, one might consider a giant impact to be merely the most extreme example of a capture scenario, or the formation of moons from the resulting debris disk to be an example of formation in a circumplanetary disk (CPD). Or the CPD material itself might be considered ``captured'' from the primordial circumstellar disk.

\vspace{0.5em}

The formation of Earth's moon is generally considered to be the result of the first scenario, a giant impact \citep[e.g.][]{Canup:2001}, as is Pluto's large moon Charon and its retinue of smaller satellites \citep{Stern:2006, Canup:2011}. The Galilean moons, found orbiting co-planar to the equator of Jupiter and in an elegant Laplace resonance, are thought to have formed instead within a circumplanetary disk, built of material stolen from the protoplanetary disk by its massive parent planet \citep[e.g.][]{Lunine:1982, Ogihara:2012, Heller:2015b, Cilibrasi:2018}. With regard to gravitational capture, the quintessential example is Neptune's largest moon Triton, with its retrograde orbit inclined significantly with respect to the ecliptic \citep{Agnor:2006}. The so-called irregular moons, sometimes on quite eccentric, inclined (possibly retrograde) orbits are defined as such because they are captured; they make up the bulk of the known moons orbiting the major planets, and the vast majority are on the order of a few kilometers in size.

\vspace{0.5em} 

At the heart of physics is the assumption / expectation that processes at work in our local corner of the Universe play out the same way elsewhere, under the same conditions. We may naturally guess then that if we see moons formed here in our Solar System, not just in great numbers but also through a variety of pathways, we ought to see an abundance of moons in other planetary systems, as well. At the very least, it is not far-fetched. At the same time, the apparent peculiarity of our Solar System when compared to the menagerie of planetary systems we have discovered in the last two decades may give us some pause. It remains an open question how common, exactly, a place like the Solar System might be in the Milky Way. But it is clear that ours is quite a bit different from systems displaying tightly packed planets \citep{TRAPPIST1, Luger:2017}, Hot Jupiters \citep[e.g.][]{Mayor:1995}, Super-Earths or Mini-Neptunes \citep[e.g.][]{Borucki:2011}, highly elliptical planets \citep[e.g.][]{Jones:2006}, and spin-orbit misalignments \citep[][]{Winn:2005, Fabrycky:2009, Huber:2013}, for example. Might exomoons then not be so common? How abundant might they be? Where should we expect to find them? How massive can they be? Might they be habitable, even (dare we say it) Earth-like?

\subsubsection{Moon System Architectures}
The most basic requirement of the moons we will occupy ourselves with here is long-term stability; if it cannot exist for a significant fraction of the planet's lifetime, we have little hope of ever detecting them after the CPD stage. We will generally observe a planetary system during its long middle age, and so an observable moon must be formed, outlast a presumably tumultuous period of infancy, and then maintain a stable orbit for billions of years. Fundamentally, we know that a moon must reside at a distance from its planet somewhere between the %\hl{Roche limit}
Roche limit (within which tidal forces will tear the moon to shreds) and the %\hl{Hill radius}
Hill radius, $R_{\mathrm{Hill}}$ (beyond which the planet's gravitational pull is no longer sufficient to hold onto the satellite, due to external forces). These are the most generous limits, however: \citealt{Domingos:2006}, for example, found that prograde moons are only stable out to 0.4895 $R_{\mathrm{Hill}}$, while retrograde moons may be stable out to 0.9309 $R_{\mathrm{Hill}}$. More recently, \citealt{RosarioFranco:2020} found a prograde stability limit out to 0.4061 $R_{\mathrm{Hill}}$, with moons out to the 0.4895 $R_{\mathrm{Hill}}$ stable in only a fraction of cases. These single values also cannot reflect the full range of critical factors in the moon's stability, such as the inclination of the satellite, the eccentricity of the planet's orbit, or the tidal quality factor, for example; the gravitational potential of the planet is generally \textit{not} spherically symmetric, and this will certainly affect stability limits.

\vspace{0.5em} 

The satellite formation pathway itself can be expected to play a significant role in the overall architecture. Consider the architectures of the satellite systems in our Solar System. In terms of mass ratios and orbital configurations, the Galilean moons look markedly different than the moons of Earth and Pluto, and these all quite a bit different from the moons of Mars, or Neptune. Exomoon architectures might therefore provide critical insights into the histories of their host planets and the planetary systems as a whole, just as the moons in our Solar System can. Every moon tells a story about the formation and evolution of its host planet.

\vspace{0.5em} 

Simulations of circumplanetary disks and their products can provide further insights into what we can expect from satellite systems formed in this way. Key factors in the growth and survival of moons include (for example): 1) the rate of inflow of material from the protoplanetary disk \citep[e.g.][]{Szulagyi:2022}, 2) timescales of gas dissipation \citep[e.g.][]{Mosqueira:2003}, 3) the dust-to-gas ratio  \citep{Canup:2002, Cilibrasi:2018}, 4) the geometry of the accretion disk or envelope \citep{Szulagyi:2016}, 5), the emergence of orbital resonances \citep{Ogihara:2012}, and 6) the temperature profile of the disk \citep{Szulagyi:2017, Fujii:2020}. The general picture is that, for moons formed in a circumplanetary disk, total satellite masses are unlikely to reach much above a mass ratio of $10^{-4}$ \citep{Canup:2006} -- entirely consistent with what we see in the Solar System. Of course, consistency with the Solar System does not imply that this is the only way it must be, so it will be important at every step to interrogate our assumptions. Future exomoon discoveries may defy our expectations.

\vspace{0.5em} 

With regard to exomoons formed by giant impact, \citealt{Malamud:2020} found that it is quite difficult to form exomoons large enough to be currently detectable in this way through a single impact event, suggesting that super-Earths will be more attractive hosts from a detection standpoiont as they will have commensurately larger moons. On the other hand, \citealt{Nakajima:2022} has recently predicted that lower mass planets ($<6 M_{\oplus}$, corresponding to $<1.6 R_{\oplus}$) will be more likely to form moons from the collision. This owes to the fact that collisions with more massive planets should result in a gas-rich disk that produces so much drag on the ejected material that it rapidly falls back down onto the planet. We may therefore want to focus detection efforts instead on smaller terrestrial planets, as these will be more likely to host ``fractionally large'' (detectable) moons. Fractionally large moons will display deeper transits with respect to the planet's transit depth, and will also result in larger amplitude transit timing variations (TTVs -- more on this later). Of course, as the giant planets of our Solar System are host to the majority of large moons (such that their extrasolar analogues are detectable by present or near-future instruments), they will also remain attractive targets in the search.

\vspace{0.5em} 

What about captured exomoons? Recent exomoon candidate claims \citep{k1625_a, k1708} are hinting at the possibility that exomoons may sometimes be far larger than any moon seen in the Solar System. These massive exomoons, if real, were hardly anticipated by the community prior to suggestions of their existence observationally, but they now have some theoretically plausible formation pathways \citep{Hamers:2018, Hansen:2019}. For now it remains difficult to predict the likelihood of these types of objects, if indeed they are real, though close encounters should theoretically occur with roughly the same frequency as planetary collisions. We will discuss these cases in more detail later on.

\subsubsection{Moon Habitability}
As we have come to know the moons in our Solar System more intimately since robotic spacecraft missions became possible, and all but the remotest worlds of the outer Solar System have been visited, we now know for certain that there is no place nearly so hospitable as the Earth to be found nearby. Still, armed with the knowledge that life can be found virtually everywhere on the Earth, even in noxious and scalding hot springs, even in the frigid, desolate abyss of the world's oceans, or at extreme depths underground, we can certainly imagine the possibility that life may have arisen, and maybe even flourishes today, in the more extreme conditions found in the moons.

\vspace{0.5em} 

Moreover, we now know that moons can host virtually all the ingredients necessary for life. Water is abundant in the Solar System, and it now widely accepted that liquid water can be found under the icy surfaces of (for example) Europa \citep[e.g.][]{Squyres:1983, Europa_ocean, Carr:1998}, Enceladus \citep{Enceladus_ocean}, Ganymede \citep{Ganymede_ocean}, Callisto \citep{Europa_ocean} and Titan \citep{Titan_ocean}. It is also evident that, while these moons of the outer Solar System lack the energetic input of the Sun we receive here on Earth, they do have internal heating, thanks to the injection of tidal energy from their host planets. This manifests as volcanoes on Io \citep{Io_volcanoes}, and geysers on Europa \citep{Roth:2014} and Enceladus \citep[e.g.][]{Hansen:2006, Porco:2006}. These geysers open up the exquisite possibility of \textit{in-situ} astrobiological experiments to search for evidence of life beneath the icy surface. Meanwhile, we find lakes of liquid hydrocarbons on Saturn's largest moon Titan \citep{Titan_lakes}, and even a thick atmosphere, which was first detected spectroscopically many decades ago \citep{Kuiper:1944}. None of these worlds come close to rivaling the clement conditions of Earth, but they nevertheless hint at the possibility that exomoons could indeed be habitable, even inhabited.

\vspace{0.5em} 

When it comes to assessing habitability, it is important to remain somewhat flexible in our thinking; we should not be so parochial as to assume life in general has all the requirements of life found here in Earth. Even so, it is worthwhile to bring our understanding of physics, chemistry, planetary science, and biology to bear on the question, knowing that which we rely on for life here on Earth, to understand the merits and shortcomings of moons as life-hosting worlds.

\vspace{0.5em} 

Important considerations include

\begin{itemize}
    \item availability of (liquid) water
    \item presence of critical elements / molecules, in adequate abundance
    \item capacity for long-term atmospheric retention
    \item degree of internal and external sources of energy (e.g. tidal heating, instellation, heat flux from young planets)
    \item magnetospheric shielding from stellar winds
    \item bombardment by charged particles along planetary magnetic field lines
    \item the effect of tidal locking (regular eclipses, long days and nights)
    \item recycling of surface material (tectonics, carbonate-sillicate cycle)
    \item orbital stability and migrational history
    \item climatic stability 
\end{itemize}

A great deal has been written about what makes a planet habitable, and the habitability of Solar System moons has elicited considerable attention, as well
\citep[e.g.][]{Reynolds:1983, Horst:2017, Vance:2018}. But significant efforts have also been made to investigate the habitability of exomoons, and whether we may assess the likelihood of their habitability without close-up observation.

\vspace{0.5em} 

The availability of water, particularly liquid water, tends to be of preeminent concern with respect to habitability. Water ice is evidently abundant in the Solar System, and the moons of the outer Solar System especially are vast reservoirs. But it is by no means guaranteed that moons will be water-rich, even in the cold environs far from a system's host star. \citealt{Heller:2015c} pointed out, for example, that radiation from hot, young planets have the potential to desiccate moons. It may be that the innermost Galilean moon, Io, was dried out in this way \citep{Io_dry}, though it remains a matter of debate. Thus, the orbital history of the moon would have to be known to fully establish habitability; a moon currently orbiting in a region suitable for liquid water could, depending on its history, have already been depleted of its water reservoir in the distant past. As we see mere snapshots in time, this history cannot be observed directly. 

\vspace{0.5em} 

Then there is the question of \textit{where} this water exists -- on the surface, or beneath it? We know from the Solar System, at least, that liquid water is possible at great distances from the star, when it is sequestered from the near-vacuum at the surface by a thick icy crust. For exomoons, \citealt{Tjoa:2020} provided a mathematical framework for predicting the possibility of subsurface oceans, finding that indeed there can be potentially habitable moons in a wide range of locations, including at great distances from the host star. Surface water, meanwhile, will depend on a variety of additional factors. %\citep[e.g.][]{Dobos:2015, Dobos:2017}.

\vspace{0.5em} 

Heating is clearly central to the question of liquid water availability, above or below the surface. But when is that heating too much? Tidal heating will play a major role \citep[e.g.][]{Scharf:2006, Dobos:2015, Forgan:2016, Dobos:2017}, as will proximity to the host star \citep[e.g.][]{Scharf:2006, Heller:2012, Heller:2013b, Hinkel:2013, Zollinger:2017}, planetary illumination \citep[e.g.][]{Heller:2013b, Forgan:2014, Dobos:2017}, and the presence of an atmosphere \citep[e.g.][]{Lammer:2014} along with its capacity for heat redistribution \citep[e.g.][]{Haqq-Misra:2018}. The picture has emerged that too much heating can result not just in the loss of volatiles, but a runaway greenhouse effect \citep[][]{Heller:2012, Hinkel:2013, Heller:2013b, Zollinger:2017}. Moons found in the planetary habitable zone may not themselves be habitable due to overheating, but those somewhat exterior to this zone, or passing it through it for a fraction of the planet's orbit, may be more hospitable \citep[][]{Hinkel:2013, Forgan:2014, Zollinger:2017}.

\vspace{0.5em} 

For surface water to exist, a substantial atmosphere must also be present. \citealt{Lammer:2014} focused on the stability of exomoon atmospheres in the planetary habitable zone, finding that low mass moons ($<0.25 M_{\oplus}$) may not be able to sustain them, due to the ablative effects of stellar winds, especially in the first $\sim100$ million years of the system. More massive moons may fare better, as can moons at greater distances from the star (Saturn's largest moon Titan clearly shows that low mass moons can sustain atmospheres in some cases). And for free-floating planets, the deleterious effects of the star go away entirely, such that liquid water again becomes plausible for any moons that may accompany them \citep{Avila:2021}.  

\vspace{0.5em} 

The presence of a magnetosphere will also play a role. If the moon is found within the magnetosphere of the planet, it could be shielded from the stellar wind in a manner that helps it retain its atmosphere, but on the other hand, high-energy particles may be directed along the field lines down toward the moons, having adverse effects \citep{Heller:2013}. If the moon also has a magnetosphere it could be protected from harmful radiation, whether it is inside or outside the planet's magnetosphere, and there may be mutual advantage for both the planet and the moon in terms of shielding \citep{Green:2021}. If exomoon atmospheres are present, they may be conceivably be detected with transit spectroscopy with JWST \citep{Kaltenegger:2010}.

\subsubsection{Ring Formation, Composition, and Architectures}
There are a variety of mechanisms that might lead to the presence of rings, namely: 1) the collision of two minor bodies, 2) tidal destruction or stripping of a minor body, 3) the incomplete dissipation of a primordial circumplanetary disk, and 4) ejection of material from a moon. Let us briefly consider each of these in turn.

\vspace{0.5em} 

Regarding a collision of two minor bodies, this could occur at any time during the lifetime of the planet, in the ancient past or comparatively recently \citep[e.g.][]{moon_collision}. Circumplanetary disks, like their larger cousins the protoplanetary disks, will see a great number of internal collisions during the early years of formation owing to a higher density of material. These collisions may result in the fracture or destruction of proto-satellites, resulting in the formation of a ring. Moon system architectures are also subject to change over time as the orbits of the moons are altered, due to mutual perturbation between two moons or as angular momentum is exchanged between the planet and the moons. The result could be the eventual collision of two satellites whose orbits were once close but not intersecting. 

\vspace{0.5em} 

Bombardment from outside the planet's gravitational sphere of influence is also likely, especially in the early years of the system's lifetime \citep[e.g.][]{late_heavy_bombardment}, but also at any time in the planet's lifetime; a major cometary impact was observed in the Jovian system in 1994, for example \citep[Shoemaker-Levy 9,][]{Shoemaker-Levy9}. Comets or asteroids formed in a different region of the planetary system may have their orbits perturbed, eventually ending up on a collision course with a planet, or being captured by a planet. This clearly has the potential to be disruptive to the circumplanetary environs and, providing the comet's size is comparable to that of a small moon, could even pulverize whatever unfortunate satellite happens to be in its way. The extensive cratering of moons throughout the Solar System are a testament to this history of bombardment.

\vspace{0.5em} 

Tidal disruption of a moon may occur when the satellite migrates inward to the Roche limit -- the point at which gravitational tides on an orbiting body are such that the differential force on a satellite's near and far sides exceed the internal strength of the body (held together by gravity or tensile force), tearing the satellite apart. The resulting material remains in orbit (providing it does not experience drag forces that bring it down to the planet's surface), and in short order is spread across the entire orbit of the satellite to form a ring. Once again, because satellite orbits can change over long timescales, moons have the potential for living comparatively long lives before meeting their cataclysmic end in this manner. The tidal destruction or stripping of a hypothetical moon or centaur has been advanced as an explanation for Saturn's extensive ring system \citep[e.g.][]{Saturn_centaur_stripping, Saturn_moon_stripping, Saturn_chrysalis}.

\vspace{0.5em} 

It is also possible that rings may be the leftovers of the disk of material out of which moons were formed. As we have discussed, it is believed that the regular satellites of the gas giants formed out of a circumplanetary disk, analogous to the formation of planets in a protoplanetary disk. Evidently, these disks dissipate in time due to a number of processes acting concurrently: either the material is incorporated into the orbiting bodies, or it falls onto the planet, or it is ejected through dynamical interactions, or it is blown away by winds from host star. Broadly speaking, the disk of material that once surrounded the early Sun, out of which the planets were formed, has long since abated. Arguably, it has not completely disappeared, however: both the asteroid belt and the Kuiper belt could be considered massive circumstellar rings that are left over from the primordial disk. Why were they not incorporated into the other planets, or able to form one of their own? In the case of the asteroid belt, the influence of Jupiter may have played a role in suppressing their accumulation \citep[e.g.][]{Jupiter_asteroids}. In the Kuiper belt, there may be a mix of processes at play, for example, scattering by the giant planets \citep[e.g.][]{scattered_disk, kuiper_belt_origin}, or simply being too spread out and too low mass to accumulate into another planet \citep[e.g.][]{kuiper_belt_mass1}. In any case, we might see these as massive analogues of a circumplanetary disk that has dissipated incompletely. Dynamical interactions between large moons and the disk material might then also have the effect of preventing the coalescing of a moon, leaving instead a ring.

\vspace{0.5em} 

Finally, consider the ejection of material from one or more moons in orbit around the host planet. We now know of several examples of volcanism or cryo-volcanism in the Solar System, on Io \citep{Io_volcanoes}, Enceladus \citep{Porco:2006}, and Europa \citep{Europa_geyser1, Europa_geyser2} %\citep[e.g.][]{Io_volcanoes, Porco:2006, Europa_geyser1, Europa_geyser2}. 
Due to the low mass of these moons, escape velocities are not exceptionally high, and so material spewed from their surfaces find their way into circumplanetary space. Once again, this material eventually populates the entire orbit of the moon. Saturn's E ring, for example, is believed to be associated with the geysers from Enceladus \citep{Enceladus_Ering}.  In other circumstances, material can be ejected from the surface through minor impacts. These impacts are responsible for rings that bear the names of the moons that have populated them: Janus and Epimetheus \citep{Janus_Epimetheus_ring}, Methone, Anthe, and Pallene \citep{Hedman:2009}, and Phoebe \citep{Phoebe_ring}. Clearly, ejecta alone cannot be responsible for the entire Saturnian ring system, but such a process could certainly explain rings around exoplanets as it has been well demonstrated process in our own Solar System.

\vspace{0.5em} 

The nature of these various mechanisms mentioned above will partly inform the inferred age and lifetimes of these rings. Let us take Saturn again as an example; if the rings are primordial, they have existed in some form for several billion years. Conversely, a collision between two objects, particularly if one object came from beyond the planet's sphere of influence, might not necessarily be particularly old. The tidal destruction of a moon could have also happened in the relatively recent past, as moons may gradually lose angular momentum over the course of billions of years. If the ring is composed of ejecta from a moon, this ring could be young or old, but not need not on its own be long-term stable, as it is continues to be fed. 

\vspace{0.5em} 

What are planetary rings made of? Here in the Solar System, they are are made mostly of ice (Saturn and Uranus) %\citep{Saturn_ring_ice} 
or dust (Jupiter), or a combination of both (Neptune). These are of course the materials that are found in abundance in the outer reaches of the Solar System where the gas giants reside. But could ring systems be composed of rocky material instead? Though we have nothing like it in the Solar System (ignoring the interpretation of the asteroid belt as a giant ring system), there is no reason a rocky ring could not exist providing it has some formation pathway. Looking back at the mechanisms we've described above, the requirement is simply a repository of rocky material in the planet's vicinity, either indigenous or introduced.

\vspace{0.5em}

If we are to find rings around an exoplanet residing in a region of high instellation, they would almost certainly not be made of ice. Icy rings would be sublimated and simply could not survive on long timescales. Were we to observe icy rings around a warm exoplanet, it's a good bet the rings would be very young. From an observational standpoint, these planets close to their host stars (whether they are rocky or gaseous), represent the most convenient targets for transit observations. Therefore, we will be especially interested in knowing whether rocky rings exist, because if we see rings around these planets, they almost certainly must be made of rock or dust.

\vspace{0.5em}

The ring of the major planets in our Solar System are found to be mostly co-planar with their planet's equator. This alignment is due to the planet's non-spherical gravitational potential associated with the internal structure of the planet, with its non-uniform, non-spheroidal mass distribution. It is then an intriguing possibility that observing ringed exoplanets may provide a means for probing the internal structures of these planets \citep[][]{Ragozzine:2009}, which are otherwise unavailable to us observationally. Ringed exoplanets may also be reveal the obliquity of the planet, which is again not directly observable otherwise.

\vspace{0.5em}

As we have seen in the Solar System (most spectacularly around Saturn), gaps may form in the rings. These gaps may be opened up either by a moon embedded within the rings, or by the presence of an outer moon in resonance with the rings \citep[see][]{Goldreich:1982}. For rings in transiting exoplanet systems, the gaps may be discernible as moments of brightening in the light curve during the transit, as was seen in observations of the system J1407 (discussed further below). Insomuch as the gaps are produced by the presence of satellites, either embedded or shepherding from outside, the gaps could be indirect evidence for the presence of moons \citep[e.g.][]{Kenworthy:2015}. However, it is worth some caution in interpreting these gaps as such. As is the case with protoplanetary disks, there may be explanations for gaps that are not associated with the presence of bodies embedded or exterior to the gap. In the case of J1407, for example, there is some indication that the gap could be due to stellar tides experienced by the system due to a significant eccentricity \citep[][]{Sutton:2019}.

\vspace{0.5em}

\subsection{Observational Signatures}

Now that we have discussed those features we anticipate from moons and rings -- either because we have directly observed them in the Solar System or because we have speculated on their plausibility -- let us now consider the observability of these features.  What can we hope to see, in the future if not at present?

\subsubsection{Signatures of Exomoons}
\label{sec:signatures_of_exomoons}
Just like planets, moons will be illuminated by the light of their host star. But, like the vast majority of exoplanets, the reflected light we receive here on Earth from exomoons will be far too feeble to be detectable in the near-term. Instead, we may turn to infrared observations, which can reveal heat signatures of these worlds. Direct imaging has revealed giant planets at great distances (several AU) from their host stars, but their detection is owed to their relative youth: they are generally so young as to still be cooling off from formation \citep[e.g.][]{Chauvin:2004, Marois:2008}, so they are bright at infrared wavelengths. When it comes to using direct imaging in the search for exomoons, we must also consider the problem of angular resolution. At the great distances to these systems, the angular distance between a planet and a moon will generally be exceedingly small. Nevertheless, with sufficient aperture size, it is conceivable that these moons could be seen one day with reflected or emitted light, and spatially resolved. 

\vspace{0.5em} 

If unresolved, we may still be able to detect a moon's presence indirectly, in the form of a flux centroid shift across different wavelengths \citep{Agol:2015}. Remarkably, it is possible that while the planet outshines the moon at some wavelengths, the moon can outshine the planet at others, owing to different thermal emission and/or reflective profiles. Thus, for the unresolved pair, the center of light on the detector will shift towards the brighter of the two objects, and this can be measured at a sub-pixel level. Tidally heated exomoons may also be discernible if the planet-moon is unresolved, through an analysis of their combined spectral energy distribution \citep{Limbach:2013, Jager:2021, Kleisioti:2023, Tokadjian:2023}.

\vspace{0.5em} 

We may instead try to observe the moon \textit{in relationship to the host planet}. This may come in a variety of forms. It has been suggested, by analogy with the Solar System, that there may be interactions between the planet's magnetosphere and the moon, producing detectable radio emissions and modulations \citep{Noyola:2014, Noyola:2016}. Or if the orientation of the system is advantageous, it could be possible to see the moon transit the face of the planet, analogous with a planetary transit of a star \citep{Cabrera:2007}. This technique has recently been explored also as an approach to detecting satellites around isolated or free-floating planetary-mass objects \citep{Limbach:2021}. For moons with significantly small semi-major axes, the geometric transit probabilities may be upwards of 10\%, offering a comparatively good chance of detecting these objects if reasonably large survey is carried out. \citealt{Limbach:2023} recently proposed the TEMPO survey, a 30-day campaign with the Nancy Grace Roman Space Telescope, to observe the Orion Nebula Cluster in search of these exosatellites. They predict a yield of $\sim10$ detections of satellites around free floating planets and another $\sim50$ objects orbiting brown dwarfs. 

\vspace{0.5em}

Then there is the gravitational influence the moon exerts on the planet. \citealt{Vanderburg:2018} has suggested that the reflex motion of a planet hosting a fractionally large moon ($\gtrsim$ 1\% mass ratio) could be detectable with direct imaging integral field spectroscopy in the near future. A proof of concept for this idea was carried out in \citealt{Vanderburg:2021} with the directly imaged planetary system HR 8799, though the constraints they are able to place on possible moon masses are currently not especially illuminating. In time, however, this approach holds great promise. Most recently, \citealt{Ruffio:2023} undertook another study, this time of the HR 7672 B system, finding again that current technology can only put very high upper limits (1-4\%) on possible companion mass ratios, but that future 30-meter class telescopes have the potential to detect mass ratios more akin to those in the Solar System.

\vspace{0.5em} 

Planets and their moons may also be detected through microlensing, and such signals may have already been identified \citep{Bennett:2014, Hwang:2018, Miyazaki:2018}. The light of a background source can be amplified dramatically by the close (sky-projected) passing of objects, even sub-Earth mass objects, and through analysis of a microlensing light curve and modeling the caustics, the masses and architectures of these passing systems may be inferred. There are some challenging degeneracies associated with this approach, however, which we will discuss later. Recently, \citealt{Bachelet:2022} simulated a forecast 40,000 moon lensing events that might occur during a joint Euclid-Roman microlensing survey, finding that 1\% of the events could be detectable. 

\vspace{0.5em} 

Finally, there is the moon's impact on the light we receive from the planet's host star. Stars produce several orders of magnitude more photons than the planets or their environs, so they are by far the easiest objects to measure and study.

\vspace{0.5em}

One intriguing suggestion is that exomoons may be detected through pulsar timing \citep{Lewis:2008, Pasqua:2014}. Readers may recall that the first exoplanets ever discovered were found via pulsar timing \citep{Wolszczan:1992}. While the gravitational influence of the planet-moon system is far too small to ever be seen in the radial velocity of the host star (the amplitude of modulation being well below the noise floor from stellar surface activity), our capacity to measure pulsar rotation periods with exquisite precision make this slight perturbation detectable through a time-of-arrival analysis. 

\begin{figure}
    \centering
    \includegraphics[width=\columnwidth]{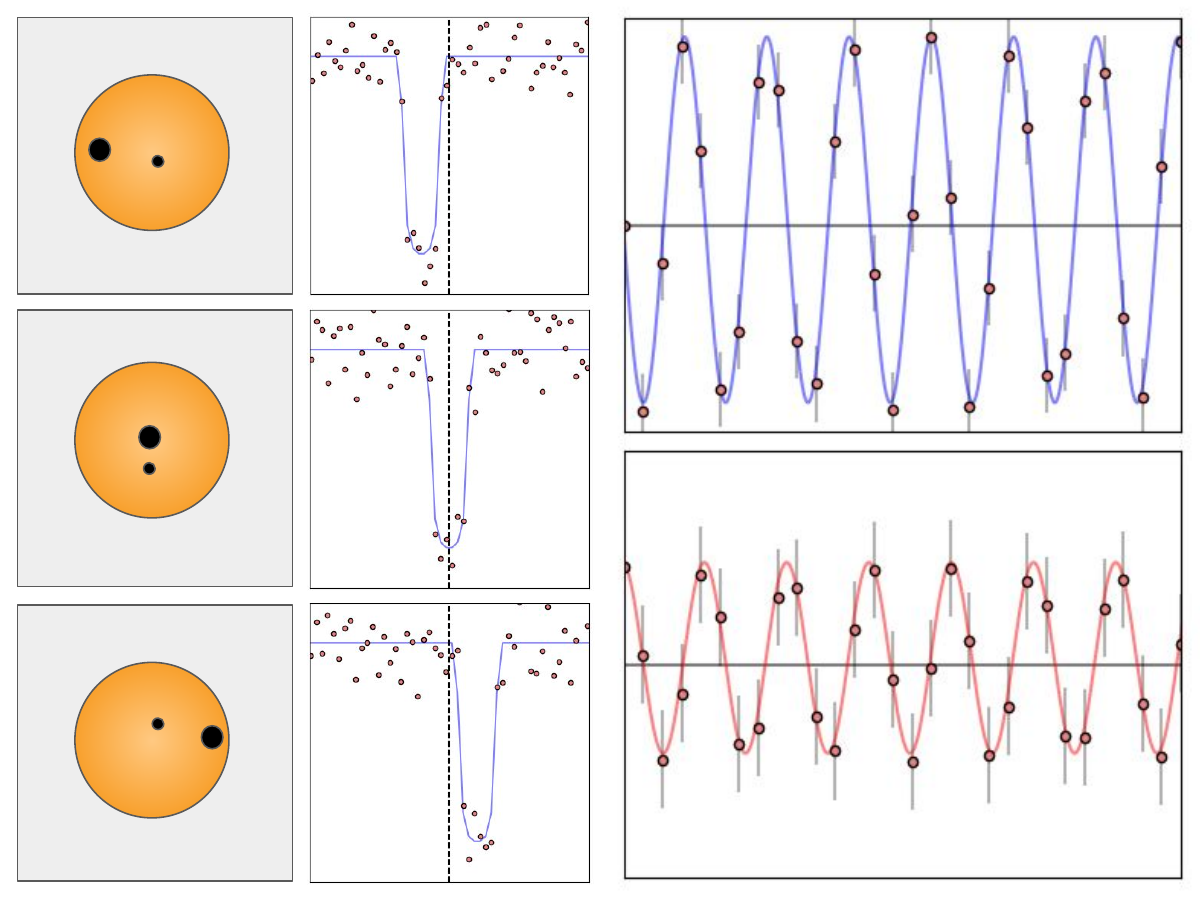}
    \caption{\textit{Left:} diagram of a planet and moon transiting in three configurations. \textit{Center:} transit time variations associated with the configurations shown at left. \textit{Right:} Possible TTV (top) and TDV (bottom) signals for this planet, showing relative amplitudes, noises, and $\pi / 2$ phase offset.}
    \label{fig:TTV_TDV_diagram}
\end{figure}

\vspace{0.5em} 

But let us now consider transits in time-domain photometric observations, arguably the most popular approach to detecting exomoons at present. Exomoons, like their host planet, can be seen transiting the face of the star, blocking out a small percentage of the light during the event. We can generally expect exomoon transits will be shallower than that of the host planet, and will be seen leading or trailing the planetary transit, or sometimes overlapping with it, in time. Moon transits will be of roughly the same duration as the planet's transit, but not exactly, owing to their motion with respect to the barycenter. The time between planet and moon transits naturally reflects their sky-projected separation, and a maximum time separation may be computed for given moon semi-major axis. 

\vspace{0.5em} 

Substantial efforts have been made to understand the particulars of exomoon transits, and to explore novel approaches to identifying them. Perturbations of the planet's motion, due to the presence of the moon, may be measured in the form of %\hl{transit timing variations}
transit timing variations \citep[TTVs,][]{Simon:2007} and %\hl{transit duration variations}
transit duration variations \citep[TDVs,][]{Kipping:2009}. moon-induced TTVs will occur because the planet is offset from the planet-moon barycenter. As a result, during transit the planet will be observed as either leading or trailing the barycenter, resulting in transits that come early or late. Plotting transit timing effects versus transit epoch reveals that both TTVs and TDVs produce a sinusoidal pattern (see Figure \ref{fig:TTV_TDV_diagram}). The amplitudes will be dictated by the mass ratio and the semi-major axis.

\vspace{0.5em}

\begin{figure}
    \centering
    \includegraphics[width=\columnwidth]{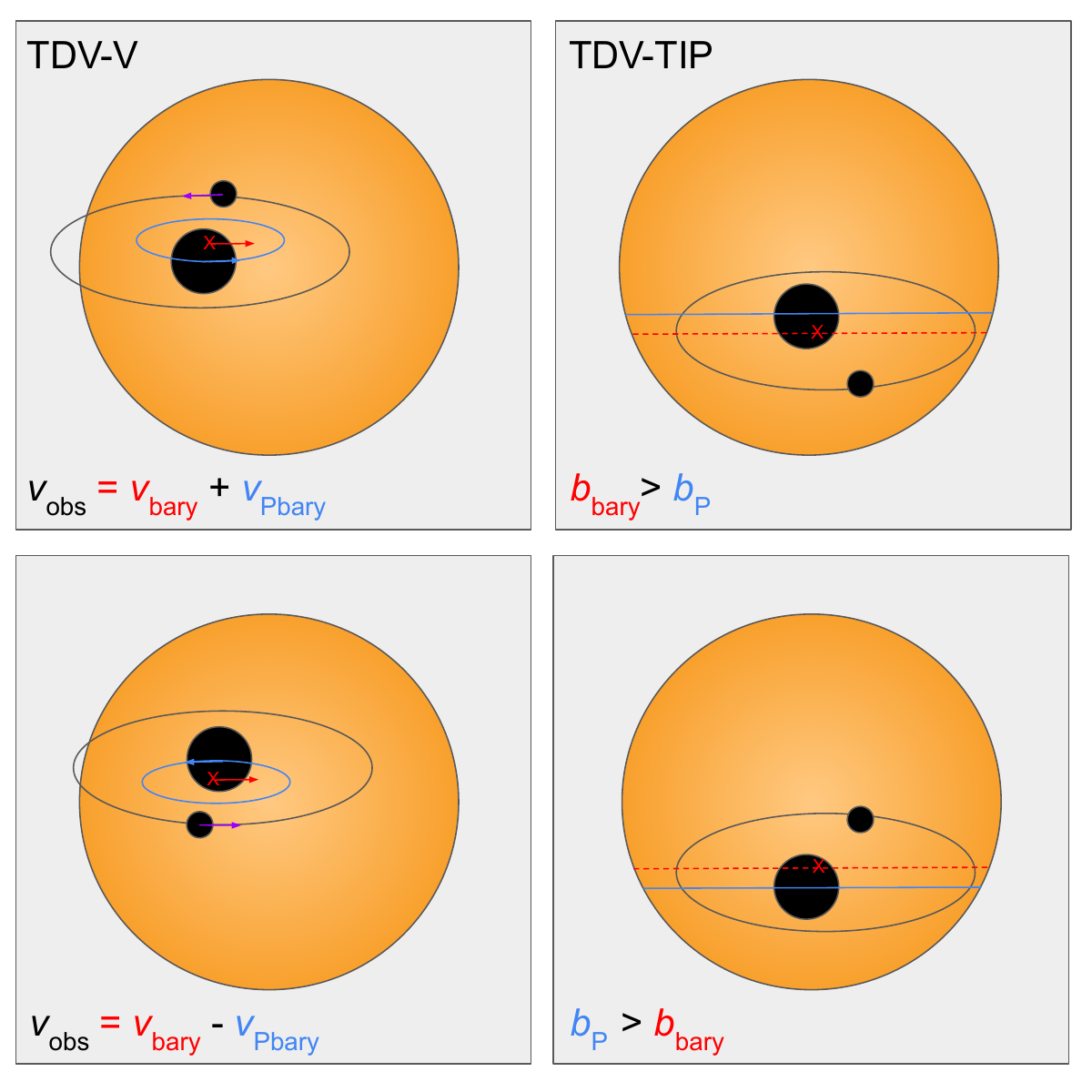}
    \caption{\textit{Left:} illustration of the geometry that results in TDV-Vs. \textit{Right:} the geometry that results in TDV-TIPs.}
    \label{fig:TDVV_TDVTIP}
\end{figure}

TDVs have two separate causes (see Figure \ref{fig:TDVV_TDVTIP}): the velocity-induced TDV \citep[TDV-V,][]{Kipping:2009} occurs because the planet has some non-zero velocity with respect to the barycenter; while the barycenter's transit duration is essentially fixed, this extra velocity component will mean the planet appears to traverse the face of the star faster or slower depending on the geometry. Meanwhile, the TDV-TIP \citep[][]{Kipping:2009b} is caused by a changing planetary transit impact parameter. For a moon with an inclined orbit, the planet will now display some vertical offset from the barycenter's impact parameter, making the transit chord longer or shorter and thus extending or shortening the duration of the transit also. Critically, moon-induced TTVs and TDVs will always be out of phase by $\pi / 2$, and this feature is unique to exomoons \citep[][]{Kipping:2009, Heller:2016b}. TDVs will typically be small, but could be substantial in the case of planets with secondaries at significant mass ratio, or binary planets \citep{Chakraborty:2023}.

\vspace{0.5em}

For moon transits coincident with the planet transits, \citealt{Rodenbeck:2020} proposed the measurement of transit radius variations (TRV) as an indicator of exomoons. In fitting a planet model to a transit that is sometimes contaminated by a moon, the effective radius of the planet will change from epoch to epoch. To employ this technique, then, observers will need to let the planet's radius in each epoch be a free parameter.

\vspace{0.5em} 

The Orbital Sampling Effect, first identified in \citealt{Heller:2014}, manifests as a distinctive removal of flux on the wings of the planetary transit in a time-averaged phase-fold of the light curve. While each individual moon transit may be lost in the noise, the cumulative effect of a moon ``carving out'' the flux before and after the planet's transit may be unmistakable given a sufficient number of transits. This technique may be used to look for small moons around an individual planet, given enough transits \citep{Heller:2014, Heller:2016}, or to interrogate the population of exomoons \citep{Hippke:2015, HEKVI}.

\vspace{0.5em} 

The astronomer interested in searching for exomoon transits now has a number of choices for photodynamical modeling. \citealt{LUNA} first provided a recipe for the \texttt{LUNA} algorithm, a code for accurately modeling exomoon transits, accounting for the complex range of possible star-planet-moon syzygys. More recently, \citealt{Pandora} produced the open-source \texttt{Pandora} code, which provides users with a fast Python code for producing photodynamical models. Shortly thereafter, \citealt{Gefera} released their own powerful modeling code \texttt{Gefera}, which supports modeling of two objects transiting a limb darkened star (including but not limited to exomoons). \citealt{Saha:2022} also published a simplified mathematical recipe for modeling exomoon transits. With these options it is now easier than ever to become involved in the search for exomoon transits.

\vspace{0.5em} 

Moving beyond monochromatic time-domain photometry could reveal additional signatures of exomoons. For example, volcanic or cryo-volcanic activity may be detectable; these may be in the form of deep transit signatures at specific wavelengths \citep{BenJaffel:2014, Oza:2019, Gebek:2020}, making the planet appear unreasonably large for their mass. Or consider the %\hl{Rossiter-McLaughlin (RM) Effect}
Rossiter-McLaughlin (RM) Effect \citep{Simon:2010, Zhuang:2012, Heller:2014b, Ruffio:2023}: any rotating star will be seen by an observer as having one blue-shifted hemisphere (rotating towards us) and one red-shifted hemisphere (rotating away from us). Assuming a planet passes in front of the star at some angle not parallel to the axis of rotation, the planet will pass over the blue- and red-shifted hemispheres in turn, resulting in a spectral shift in time (as blue light is blocked out, the spectrum is reddened, and vice versa). In this way we can infer the spin-orbit (mis)alignment of the planet with respect to its star. If the planet has a moon, then, we may see multiple shifts, as the two bodies individually cross the stellar hemispheres at different times. Observing the RM effect is certainly more demanding on time and resources than ordinary transit photometry, but can yield additional insights into the system's geometry and architecture.

\vspace{0.5em} 

Multi-band observations can also pay dividends in terms of helping to disentangle moon-like signals, which should be achromatic, from stellar surface activity, which will be color-dependent \citep[e.g.][]{Herrero:2016, Gordon:2020}. Of course, many instruments are incapable of providing simultaneous imaging in multiple bands, but a grism might be used, as in \citealt{k1625_a}. Unfortunately, probing multiple colors by segmenting the broadband signal as performed in that work results in greater flux uncertainties, but the signal may still be tested for color-dependence.

\vspace{0.5em}

\subsubsection{Signatures of Exorings}
Compared to exomoons, there are relatively fewer avenues available to us for detection of exorings, as rings will not (for example) produce detectable gravitational perturbations on the host planet. Until now, the focus of exoring detection studies has been on time-domain photometry and spectroscopy, along with direct imaging.

\vspace{0.5em}

Modeling of exoring transit signatures is complex. In addition to the planet model parameters, an exoring model must also account for 1) the radial extent of the rings, 2) the obliquity of the planet (the inclination of the rings), 3) the possible presence of gaps, 4) the refractory properties of the rings, and 5) optical depth of the rings. In view of these challenges, it is worthwhile to consider first those observable signatures that might lead us to suspect the presence of rings before going through the trouble of modeling them.

\vspace{0.5em}

A planet with rings, assuming they are not seen perfectly edge-on, can block out substantially more light than a ring-free planet, resulting in significantly deeper transits \citep{Zuluaga:2015}. Directly imaged planets may likewise be measured to have anomalously large radii \citep{Arnold:2004}. Under a ring-free assumption for the planet, this large radius when paired with a mass measurement from radial velocities could suggest the presence of an extremely low-density planet. It is worth noting that at the time of this writing, more than 200 exceptionally low-density ($\rho \lesssim 0.3$ gm cm$^{-3}$) planets have been reported, a class of objects collectively referred to as ``super-puffs''. Explanations for these inflated worlds have been put forth, including tidal heating \citep{Millholland:2019}, hydrogen-rich atmospheres that are opaque even at very low pressures \citep[e.g.][]{Lammer:2016}, and photochemical hazes \citep[][]{Gao:2020}. Even so, questions remain, and it is conceivable that some of these planets may not be inflated at all, but could instead be hosting ring systems \citep{Piro:2020}

\vspace{0.5em}

Planets with exceptionally deep transits may also be erroneously classified as eclipsing binaries (EBs) or false positives \citep{Zuluaga:2015}. In contrast to exoplanets, the majority of which have transit depths less than 0.1\%, eclipsing binaries typically show transit signals that are on the order of a few percent. They are quite common and readily encountered in large time-domain datasets like \textit{Kepler} and TESS, and for the exoplanet astronomer they may be something of a nuisance, triggering transit search algorithms and wasting time on the way towards finding genuine planets. However, the possibility of ringed planets should give exoplanet astronomers some pause before dismissing these objects on visual inspection alone.

\vspace{0.5em}

\begin{figure}
    \centering
    \includegraphics[width=\columnwidth]{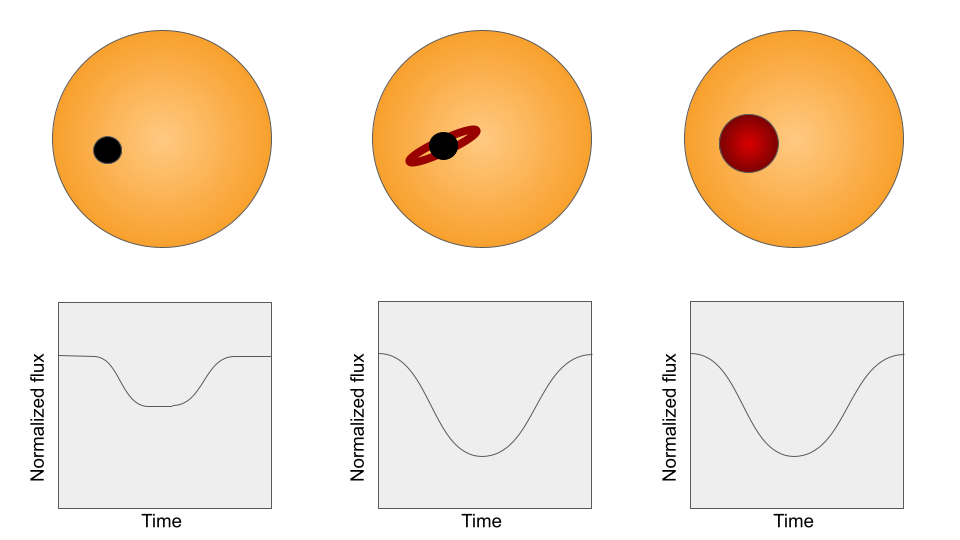}
    \caption{Planets hosting rings may be interpreted as having anomalously large radii, which could lead to their classification as a false positive, perhaps an eclipsing binary.}
    \label{fig:rings_as_EBs}
\end{figure}

In addition to deeper transits, we can expect morphological changes in the shape of the transit, particularly during ingress and egress \citep{Barnes:2004}. Again, these morphological changes could mimic an EB or other false positive scenario, with transit shapes that differ from that of an ordinary ring-free planet. Such changes to the planet's transit profile might also lead to erroneous impact parameter and / or semi-major axis solutions, as the planet's transit appears to be more grazing than it actually is, or longer in duration \citep{Zuluaga:2015}. Eccentricity solutions may also be impacted, as the planet's transit duration is extended by the much wider shadow, suggesting an eccentric orbit. Limb darkening coefficients are also vulnerable to erroneous solutions, as these are derived from the shape of planetary ingress and egress. And stellar densities, which may be derived from the planet's transit duration, may also be in error \citep{Zuluaga:2015}. Fortunately, we may check some of these model solutions with alternative methods; the planet's eccentricity, for example, can be determined independently through radial velocity measurements; a mass measurement may also rule out the eclipsing binary explanation. Stellar densities may be inferred independently from spectral analysis of the star, and limb darkening solutions may be checked against a theoretical expectation for the spectral type and observing band. Discrepant model solutions may be indicative that a moonless, ring-free planet model does not fit the data.

\vspace{0.5em}

Rings may be wide or narrow, and optically thin or thick at different wavelengths. They may also scatter light in the direction of the observer, creating a small brightening (on the order of 10s of ppm) before ingress and after egress \citep[][]{Barnes:2004, Sucerquia:2020, Zuluaga:2022}. It is even possible that the composition of the rings (rocky or icy) may be discernible based on the refractory properties of the material. Scattered light from rings may even be detectable in cases where the planet comes close to but does not actually transit \citep{Zuluaga:2022}.

\vspace{0.5em}

Targeted, time-domain spectroscopic observations of individual systems may yet yield more information leading to a ringed planet hypothesis. Like planets hosting moons, ringed planets may also be observable through the Rossiter-McLaughlin effect \citep{Ohta:2009, deMooij:2017}. Generally, the rings will display line profile morphologies that diverge from what would be produced by a ring-free planet. There may also be chromatic occultation effects due to the refractory nature of the rings. Perhaps most intriguingly, rings can have the effect of flattening out transmission spectra \citep[][]{Ohno:2022}, which are typically employed to study exoplanet atmospheres. This flattening may be responsible for some yet-unexplained flattening seen in some exoplanets' transmission spectra.

\vspace{0.5em}

Fortunately, unlike the case of moons, the transit morphology of a ringed planet is likely to stay consistent from epoch to epoch, at least on short timescales, allowing for transit stacking. Precession of the planet's spin axis could occur on long, but observable timescales \citep[that is, years-to-decades,][]{Carter:2010}. Planets orbiting close to their host star, or those on significantly elliptical orbits, may also have significant perturbation in the rings that may alter the morphology substantially from epoch to epoch \citep[][]{Sucerquia:2017, Sutton:2019}.

%% file: detection.tex
\label{sec:detection}

\section{Detection Efforts}
Having now established that there are a range of tools available to the astronomer searching for exomoons and exorings, let us now examine the efforts that have already been undertaken to find them. As before, we will deal first with exomoon search efforts before moving on to discuss the search for exorings.

\vspace{0.5em} 

Exomoons have long been predicted to be detectable with current time-domain photometry capabilities \citep{Sartoretti:1999, Szabo:2006}. The advent of major exoplanet survey missions such as \textit{Kepler} and TESS have enabled such searches, as it remains something of a numbers game; there will be many systems for which the stars are simply too faint to detect a moon above the noise, but look at enough systems and there is some hope. Photometric precision is such that we may detect sub-Earth radius worlds in transit under favorable conditions \citep[e.g.][]{Barclay:2013, Campante:2015}, and transit timings may be measured with sufficient precision to detect divergences from linear ephemeris caused by moons orders of magnitude less massive than their host planet \citep[c.f.][]{Carter:2008}. TDVs, meanwhile, are just at the limit of being detectable; by and large they are out of reach with current instruments.

\subsection{Exomoon Transit Surveys and Strategies}
Arguably the first systematic approach to detecting exomoon transits in a large survey was the Hunt for Exomoons with Kepler (HEK) project \citep[][]{HEKI, HEKII, HEKIII, HEKIV, HEKV}. In all the HEK team surveyed roughly 60 planets in search of exomoon signatures with photodynamical modeling, employing a battery of tests to test planet+moon models against planet-only models. Three strategies were employed for selection of these planets: they either 1) had intrinsic properties favorable to the presence of a detectable exomoon, 2) displayed transit morphologies suggestive of an exomoon transit, or 3) were targets of broader interest to the community. However extensive, these efforts did not yield any exceptionally promising exomoon candidates until the sixth paper in the series \citep[HEK VI,][]{HEKVI}. In that work, the authors analyzed 284 planets in search of the orbital sampling effect \citep{Heller:2014, Heller:2016} (see section \ref{sec:signatures_of_exomoons}), an effort that was also independently undertaken by \citealt{Hippke:2015}. From HEK VI emerged a single candidate, Kepler-1625 b-i, which we will discuss later. 

\vspace{0.5em} 

Most recently, \citealt{k1708} undertook a detailed investigation of 70 ``cool giant'' exoplanets, performing full photodynamical model fits to each system. Having long suspected that planets at larger distances from their host star are the most likely place for exomoons to reside (by analogy with the Solar System, and in view of both theoretical and observational results \citealt{Namouni:2010, Spalding:2016, HEKVI, Dobos:2021}), the team specifically targeted planets that had either 1) periods beyond 400 days, 2) equilibrium temperature less than 300 K, or 3) instellation below that of the Earth. This search yielded one new exomoon candidate, Kepler-1708 b-i, which we will discuss further in the following pages.

\vspace{0.5em} 
Considering the substantial computational demands of full photodynamical fits to individual planets, there has long been a desire to identify strategies for identifying signals of interest at lower cost, something akin to the Box Least Squares method \citep{BLS} that has been used extensively in search for candidate planet transit signals. The challenge is that exomoon signals are not only small and sometimes lost in the noise, they also do not show the regular periodicities of transiting planets by themselves; as we've said, the transits may show up well before, or well after, the planet's transit. This situation does not lend itself easily to phase-folding and stacking to obtain a clearer signal. Recently, however, \citealt{transit_origami} introduced ``transit origami'', a new approach to stacking exomoon transits, analogous to phase-folding a light curve on the planet's orbital period to coherently stack planet transits and recover a more coherent (convincing) signal. Transit origami attempts to solve the phase-folding problem by leveraging the transit timing variations of the planet, predicting the sky-projected planet-moon distance for each epoch, shifting the time series and stacking based on the expected position of the moon. 

\vspace{0.5em} 

There remains the problem of exomoon transits ``contaminating'' the planet's transit; they can be coinicident in time, complicating the search. \citealt{Heller:2022} have calculated that, depending on the architecture, several tens of percent of moon transits will coincide with the planet transits, with inner moons showing the greatest amount of contamination. As an example, a mere 14\% of Io's transits would be seen completely outside of the planet's transit. This increases for greater moon semi-major axes, but only up to 73\% for Jupiter's outermost large moon, Callisto. Thus, methods such as transit origami and the orbital sampling effect, which seek signals exclusively outside of the planet's transit, are throwing away a non-negligible fraction of the information content; put another way, the sensitivity of the search has been reduced considerably. A search for TRVs \citep{Rodenbeck:2020} would not be affected by this issue.

\vspace{0.5em} 

Machine learning continues to grow in popularity with the astronomical community, and these methods have recently also been brought to bear on the problem of identifying exomoon transits. In particular, convolutional neural networks (CNNs), having been demonstrated to be a powerful tool for identifying and vetting planetary transit signals \citep[e.g.][]{Shallue:2018}, have now been employed also in the exomoon transit search. \citealt{Alshehhi:2020} was the first team to build a CNN to show that exomoons can be detected in simulated data. Having undertaken this effort concurrently, \citealt{CNNs} built an ensemble of CNNs to search for exomoons, and applied their tool to 1880 Kepler Objects of Interest (KOIs). However thorough, that effort did not emerge with any exceptionally promising new candidates.

\vspace{0.5em}

\subsection{Exomoon Timing Effect Surveys and Strategies}
As we've mentioned, the gravitational influences exerted by a moon on its host planet can be imprinted in the light curve in the form of TTVs and TDVs. TTVs in particular are of significant interest to the exoplanet community, in part, because they have also played an important role in computing dynamical masses for mutually perturbing planets \citep[e.g.][]{Hadden:2017}. These TTVs may even reveal the presence of yet-unknown planets in the system, and provide constraints on the mass and location of these planets. Moreover, timing effects are relatively straightforward to search for and to measure, making them a powerful tool at the exomoon hunter's disposal. When combined with transits, together they represent three independent yet related signals that may be measured within a single data source (the light curve), and which ought to be self-consistent if the moon hypothesis holds water. 

\vspace{0.5em}

A number of studies have been aimed at investigating and exploiting these timing phenomena in the exomoon search. \citealt{Fox:2021} for example recently sought to identify candidate exomoons by searching for TTV signals consistent with the presence of a moon, and announced eight systems (out of 13 examined) they considered to be new exomoon candidates. Notably, the team explicitly set out to investigate planets for which an exomoon transit would likely be too shallow to detect, meaning that any candidates emerging from their campaign would be difficult to confirm with follow-up observations. In any case, these candidates were called into question, as \citealt{Kipping:2020} found the six candidates published in preprint each failed 2-3 tests of statistical significance, indicating weak or no evidence for moon-induced TTVs, while \citealt{Quarles:2020} found that the moon candidates were doubtful on stability grounds.

\vspace{0.5em}

Yet there remain ways forward in terms of using timing effects to interrogate the population of transiting planets in search of moons. An important consideration is whether the TTV amplitudes we measure can be attributable to a moon, or not. \citealt{Kipping:2020b} noted that some observed TTV amplitudes are be too large to be explained by a moon; a maximum TTV amplitude can be computed from the limit of a moon-planet mass ratio of unity and a satellite orbiting at the Hill radius. Thus, any system with a TTV amplitude in excess of this value must have some other gravitational influence at work (though moons may still be present).

\vspace{0.5em}

Of course, we do not only want tools that tell us when a TTV is likely \textit{not} due to an exomoon; we also want tools that help us identify targets of interest. To this end, \citealt{Kipping:2021} recently identified the so-called ``exomoon corridor'', which may help to discriminate between moon-induced TTVs and those caused by planet-planet perturbations. It so happens that moon-induced TTVs tend to display high frequency oscillations, a ``pile-up'' at short periods with fully 50\% having periods of 2-4 epochs. This evidently stands in contrast to TTVs resulting from planet-planet perturbations \citep{Hadden:2017}, which are distributed at substantially lower frequencies. Subsequently, \citealt{multimoons} found that the exomoon corridor finding holds in the more general case for systems with more than one moon, and observed the pile-up in the TTVs from \textit{Kepler} published by \citealt{Holczer:2016}. \citealt{Kipping_Yahalomi:2023} independently investigated the exomoon corridor in the \textit{Kepler} data, finding 5-10 systems that appeared most promising for exomoons, but only one (Kepler-1513 b) that survived the complete battery of tests.

\subsection{Exoring Surveys and Strategies}
Several teams have carried out searches for exoplanets hosting rings. Altogether more than 100 planets have been examined, and though we do not yet have any exceptionally promising exoring candidates, the surveys and the individual objects they scrutinized deserve brief mention here.

\vspace{0.5em}

\citealt{Heising:2015} was the first to carry out an investigation of multiple planets in search of exorings: a total of 21 Hot Jupiter's found in the \textit{Kepler} data. As such, these planets are all extremely different from the ringed planets of the outer Solar System, and icy rings would be out of the question, but rocky or dusty rings could be possible. None of these 21 planets showed evidence of rings, but the \textit{Kepler} photometry was found to be sufficient to rule out Saturn-like rings in 12 of them. Their modeling also indicated that rings may be detectable even at low inclinations, that is, seen only 5-10 degrees from edge-on. Meanwhile, \citealt{Santos:2015} also considered the possibility of rings around Hot Jupiters, performing an analysis of the hot Jupiter 51 Pegasi b which, based on a phase curve of reflected light, shows some evidence of being inflated \citep{Martins:2015}. Ultimately, \citealt{Santos:2015} found that while tilted rings could conceivably explain the observations, this architecture is unlikely from a stability standpoint. Meanwhile, rings co-planar to the planet's orbit would have insufficient incident flux to explain the observations.

\vspace{0.5em}

Not long after, \citealt{Aizawa:2017} conducted a search of some 89 long-period planets in the \textit{Kepler} data and emerged with a single candidate exoring: KIC10403228. Extensive modeling was performed on this object, which displayed only a single, V-shaped transit in the \textit{Kepler} data, and though the ring model performed well, the authors could not rule out a circumplanetary disk or a hierarchical triple system as alternative explanations. There appears to have been no subsequent follow-up of this target. That same year, \citealt{CoRoT-9b} examined two transits of CoRoT-9b observed by \textit{Spitzer} in search of both exomoons and exorings but both were excluded by the data. This planet was chosen because of its advantageous physical properties (high mass, long-period planet). Similarly, \citealt{Hatchett:2018} examined KOI-422.01, again for its long period and large radius, but found no evidence that a ring model outperforms a ring-free planet model fit.

\subsection{Notable Exomoon and Exoring Cases}
To date there have been several systems of interest published in the literature, possible exomoons or exoring systems detected directly or indirectly through a variety of techniques. Let us now briefly examine these targets.

\vspace{0.5em}

But first, a few words on nomenclature. As these systems suggest, there is sometimes a blurry line between planet and moon (and for that matter, between a ring and a disk). For example, supposing an unbound object is several times the mass of Jupiter and hosts a satellite, but a precise mass for the primary is unavailable; is this a rogue Jupiter with a moon, or a brown dwarf with a planet? What are we even supposed to call the satellite of a brown dwarf? Perhaps ``brown dwarf satellite'' is appropriate enough. ``Companion'' is perhaps sufficiently vague as to accommodate all possible interpretations. But there seems to be a particular excitement associated with the word ``exomoon'', so researchers may naturally be tempted to use it even in edge-cases. 

\vspace{0.5em}

Regarding what makes an object a ``planet'', some planetary scientists are evidently chiefly concerned with geophysical properties, or historical terminology \citep[e.g.][]{Metzger:2022}, and are comparatively less concerned with formation pathways. To them, all moons might be called planets. Astronomers, meanwhile, may care more about the origin of these objects; supposing an object is observed in isolation from a host star and has a companion -- was it formed in isolation (like a star), or did it form in a circumstellar disk (like a planet), and get ejected through a gravitational encounter with another massive object? The dividing line between planet and brown dwarf is often considered to be the deuterium burning limit, around 13 Jupiter masses \citep[see][]{Spiegel:2011}, but does that cleanly separate, say, planets that ate too much to be proper planets from stars that ate too little to be proper stars? 

\vspace{0.5em}

To ere on the side of generosity in our classification, if the community has referred to these objects as exomoon candidates, we will treat them as such, and we will not let terminology box us in too much with regard to considering the diversity of objects we find in the Universe, some of which have defied our expectations.

\vspace{0.5em}
% https://ui.adsabs.harvard.edu/abs/2015ApJ...800..126K/abstract
\noindent\textbf{J1407 b (2012).}
Arguably the first evidence for the presence of exorings and possibly exomoons came in 2012 with the discovery of J1407 b, a massive object seen transiting a pre-main sequence star with what was, apparently, a large disk of material surrounding it. This event lasted a remarkable 56 days, suggesting a truly massive disk or ring system. Intriguingly, this disk displayed gaps, which might be attributable to the presence of one or more moons. \citealt{Kenworthy:2015} provided further analysis of this system, finding a best fitting model consisting of 37 rings extending up to 0.6 AU from the host object. A 0.8 $M_{\oplus}$ moon was proposed to explain a significant gap in the rings seen at 0.4 AU from the primary. \citealt{Kenworthy:2015b} placed additional constraints on the mass and period of the object, finding a period range between 3.5 and 13.8 years, and a $3 \sigma$ upper-limit for the mass of 80 $M_{\mathrm{Jup}}$. Indications were that the object is on an eccentric orbit, no less than $e = 0.3$, but more likely with $e > 0.6$, raising some question as to the longevity of the observed disk. \citealt{Rieder:2016} performed $N-$body simulations to examine this question, strongly ruling out the possibility of a prograde ring system, but otherwise finding a solution in line with previous estimates (60 $M_{\mathrm{Jup}} < M < 100 M_{\mathrm{Jup}}$, $P \approx 11$ years) that could survive for at least $10^4$ orbits.

\vspace{0.5em}

However, \citealt{Mentel:2018} has found through an analysis of archival data dating back to 1890 that a substantial fraction of the allowable periods from earlier estimates are ruled out, as no transits were seen. While a bound solution is still compatible with the data, it was suggested that J1407 b may not be bound to the star at all and was instead a chance passing. \citealt{Sutton:2019} found that external moons in mean-motion resonance could not explain the gaps seen in the rings, but that embedded moons could be responsible. \citealt{Speedie:2020} analyzed the effect of an disk inclined with respect to the orbit of the planet (as it would need to be to be responsible for substantial dimming), and found that such a geometry significantly limits the radial extent of rings. Moreover, the perturbations associated with an inclined disk could open up gaps, thus potentially obviating the need for a moon to produce them.

\vspace{0.5em}

Recent observations with ALMA reported in \citealt{Kenworthy:2020} now lend additional credence to the idea that the event was a chance passing, finding evidence for an object at the predicted distance from J1407 based on a inferred velocity that would require the object to be unbound. Being unbound would of course relieve any tension with the size of the disk, its proximity to the host star and its eccentric orbit. Additional observations with ALMA could confirm this scenario, or show that the ALMA detection is merely a background galaxy, in which case, additional efforts to see J1407 b transit again could be warranted. \citep{Sutton:2022} performed a new analysis of the system's stability, finding that moons were unlikely to form in the case of a system with a significant eccentricity, but that the observed gap could be explained by stellar tides.

\vspace{0.5em}

The J1407 b detection is thus the first in a  growing number of intriguing but ultimately frustrating exomoon candidates, defying our desire for fast and unambiguous confirmation.

\vspace{1.0em}
\vspace{1.0em}

\noindent \textbf{MOA-2011-BLG-262 (2014).} In 2014 we were introduced to the system known as MOA-2011-BLG-262 detected through microlensing \citep{Bennett:2014}. The team labeled the object a ``free-floating exoplanet-exomoon system'', as there were only two objects detected. The finding was particularly exciting, because it demonstrated for the first time the capability of microlensing surveys to identify low mass objects, particularly when they are companions to more massive objects. In short, they are sensitive to moons. The team found a best fitting solution of a planet a few times the mass of Jupiter with a sub-Earth mass satellite. 

\vspace{0.5em}

However, there are some difficulties with the results. The planet-moon solution requires the lens to be relatively close to the Earth; if it is more distant, the masses will be scaled up, in which case the system is more likely to be a stellar mass primary with a brown dwarf mass secondary. The timescale of the event becomes relevant, and lends credence to the exomoon hypothesis, as the short duration of the event implies a high proper motion for the lens, and high proper motions are generally correlated with nearby systems. On the other hand, higher mass objects will have larger Einstein radii, which makes lensing events correspondingly more likely. At the same time, a lower proper motion solution could not be excluded. In the end, there remains two competing and roughly equally probable model solutions, so the exoplanet-exomoon system, particularly in light of its novelty, could not be definitively claimed. Unfortunately, the nature of microlensing events is such that the lensing objects can generally only be seen directly once; only the lensed background light source may be studied in greater detail after the event.

\vspace{1.0em}

\noindent \textbf{OGLE-2015-BLG-1459L (2015).} Another important case study in the search for exomoons via microlensing comes to us from \citealt{Hwang:2018}, who identified a second, intriguing microlensing event. In contrast to the free-floating objects found in \citealt{Bennett:2014}, this time the data suggested the possibility of a bound planet+moon in orbit around a host star. 

\vspace{0.5em}

However, the authors highlighted a critical degeneracy that makes characterising such systems with microlensing difficult, and ultimately found an exomoon is not the favored solution. The event clearly indicates the presence of 4 objects, but these 4 objects could have been 1) three lenses and one light source (3L1S: host star, planet, and moon), 2) two lenses and two sources (2L2S: brown dwarf and Neptune-mass planet lensing a binary), or 3) one lens and three sources (1L3S: a planet and a background hierarchical triple star system). The degeneracy is resolved by incorporating color information; multiple objects lensing a single light source will show no evidence of chromatic effects during the event, as each object is amplifying the light from the same star. On the other hand, a single object lensing multiple sources will amplify the (presumably different color) light sources to different degrees during the event, and this can be discerned if contemporaneous color information can be incorporated. Bringing in data from MOA thus helped resolve the degeneracy in favor of the one lens, three source model.

\vspace{0.5em}

As such, OGLE-2015-BLG-1459L has generally not been considered a likely exomoon candidate. Nevertheless, the result is important to keep in mind, as the authors argue that these sorts of degeneracies will be not be resolvable by the Roman Space Telescope (formerly WFIRST) unless a higher cadence (at least 9 observations per minute) and near-simultaneous observations in a second bandpass are available, for color information.

\vspace{1.0em}

\noindent \textbf{Kepler-1625 b (2017).} In the course of searching for the orbital sampling effect \citep{Heller:2014} in the \textit{Kepler} data, \citealt{HEKVI} identified Kepler-1625 b as a potential exomoon host, based on the morphology of three transits suggesting the presence of a large moon. The team's analysis showed that an exomoon explained the data well, but another transit observation with the Hubble Space Telescope (HST) would be required to confirm the presence of the moon. \citealt{Rodenbeck:2018} performed an independent analysis of the \textit{Kepler} data, finding formal evidence in favor of the moon but questioning the robustness of the detection. 

\vspace{0.5em}

The first (and so far only) observation of an exomoon candidate with HST was carried out in October 2017, observing a transit of Kepler-1625 b for approximately 40 hours. Following a joint analysis of the HST and \textit{Kepler} data, \citealt{k1625_a} claimed evidence for a moon but stopped short of declaring it a confirmed detection. The claim was based primarily on 1) the apparent presence of a moon transit following the planet's transit, and 2) a $\sim22$ minute amplitude TTV which could not be attributed to another planet in the system (as no other planets have been detected so far).

\vspace{0.5em}

Independent analyses of the joint \textit{Kepler}-HST data by two separate teams called the detection into question \citep{Heller:2019, Kreidberg:2019}. Both teams confirmed the presence of the TTV, but only one team \citep{Heller:2019} saw evidence for the post-planetary transit flux dip attributed to the moon; the other team \citep{Kreidberg:2019} did not. \citealt{Heller:2019} suggested the possibility of a non-transiting planet hot Jupiter to explain the TTV, a hypothesis that could be tested with radial velocity observations of the system. The first such study from \citealt{Timmermann:2020} validated the planetary mass of Kepler-1625 b but was not sensitive to the presence of any other planets in the system. The discovery team subsequently published a follow-up analysis \citep{k1625_b}, addressing additional unanswered questions regarding the moon candidate and ultimately maintaining the system remains a candidate.

\vspace{0.5em}

Kepler-1625 b-i, as it came to be called -- the Roman numeral used following the long established convention for naming moons, and lower-case to indicate its candidate status; this convention has not yet been officially established -- elicited a number of theory papers attempting to explain its formation, as its large size (comparable to that of Neptune) naturally raised questions as to whether something like it could even exist. \citealt{Heller_K1625_formation} was the first to examine the issue, concluding that its formation cannot be explained by formation in a circumplanetary disk, and that capture during a binary encounter is possible, but difficult. Further analyses suggested alternative capture scenarios, including tidal capture \citep{Hamers:2018} and pull-down capture \citep{Hansen:2019}. \citealt{Moraes:2020} found that an in-situ formation for the moon is theoretically possible, though it requires a particularly massive circumplanetary disk.

\vspace{0.5em}

The system's largely unanticipated nature prompted still other investigations, particularly with regard to the possibility of other moons in the system. \citealt{submoons} were the first to study whether moons might be able to sustain their own moons, an idea further explored by \citealt{RosarioFranco:2020} in relation to Kepler-1625 b. The possibility of additional moons in the system was explored in \citealt{Moraes:2022}, finding that a second, smaller moon could be stable in some configurations. \citealt{cronomoons} examined the possibility that, just as ringed planets may be masquerading as inflated planets, the inferred large radius for the Kepler-1625 b-i may in fact be due to a ring around the moon. Such a ring could be stable with the moon's inferred orbital characteristics.

\vspace{1.0em}

\noindent \textbf{MOA-2015-BLG-337 (2018).} Another borderline microlensing case comes to us from \citealt{Miyazaki:2018}, who studied an event they attributed to either a) a brown dwarf / planetary mass primary with a super-Neptune satellite (mass ratio $\sim 10^{-2}$), or b) a brown dwarf binary (mass ratio $\sim 10^{-1}$). The team found the former model was slightly favored, but not to such a degree that the brown dwarf binary model could be ruled out. Once again, even the more ``moon-like'' of these two models raise the question of when we ought to be calling this -- is it a planet and a moon or a brown dwarf and a... satellite? Interestingly, the team never refers to the lower mass object as a moon or a satellite, preferring instead the word ``companion.'' Indeed, this framework seems equally applicable to a planet-moon case as to a binary brown dwarf case, so perhaps it is the most appropriate choice in these cases. 

\vspace{1.0em}

\noindent \textbf{WASP-49 b (2019).} It has been suggested that small, volcanically active exomoons (``exo-Ios'') may be detected indirectly by detecting gases seen in absorption during the planet's transit. In this scenario, there may be a torus of material around the host planet \citep{Johnson:2006}, presenting signatures suggesting the presence of gases at extremely high altitudes (well in excess of the planet's inferred radius from the broadband transit depth). This has been imaged in the vicinity of Jupiter \citep[e.g.][]{Wilson:2002}, so it is certainly plausible that it may be seen in other systems, as well. \citealt{Oza:2019} studied 14 Hot Jupiter systems in search of these signatures, and identified one planet, WASP-49 b, which they suggested had signatures consistent with the presence of an exo-Io. In particular, deep sodium (NA I) absorption is observed for the system, which the authors argue could be attributable to outgassing from a moon.

\vspace{0.5em}

Critical to this question is whether a moon of this sort, orbiting so close to the parent star, can be stable. \citealt{Oza:2019} argue that (following \citealt{Cassidy:2009}), tidal quality factors $Q$ may be sufficiently high that stable moons are possible in this extreme environment. Even so, the stability regime is exceptionally narrow: for WASP-49 b, the authors computed a Roche limit of 1.19 $R_P$ and a stability limit out to only 1.74 $R_P$. For comparison, the innermost moon of Jupiter is the $\sim$ 20 km wide Metis, orbiting around 1.8 $R_J$, and Io is found around 6 $R_J$. \citealt{Gebek:2020} studied the case of WASP-49 b further, and concluded that while the moon hypothesis remains possible, other models also fit the data. \citealt{Kawauchi:2022} has attributed the absorption feature to a high-altitude thermosphere and atmospheric escape. The authors urge follow-up observations ot further investigate the exomoon hypothesis, and to search for more such objects. \citealt{Narang:2023a} undertook a search for radio emission from this system, but were unsuccessful in recovering a signal. The authors attribute this non-detection (in the event that a moon is in fact present) to time varying signal strengths, an overestimation of the cyclotron frequency, or overestimation of the flux density. A similar search, this time around WASP-69 b (another planet studied by \citealt{Oza:2019}), was carried out by \citealt{Narang:2023b} but also resulted in a non-detection.

\vspace{1.0em}

\noindent \textbf{PDS 70 c (2021).} We turn now to consider PDS 70, a young (5.4 million year old) T Tauri star surrounded by a large protoplanetary disk, first identified with IRAS by \citealt{PDS70}. Evidence for planet formation was first reported in \citealt{Riaud:2006} using the NAOS-CONICA instrument on VLT, and a substantial gap was finally resolved with the HiCIAO instrument on Subaru \citep{Hashimoto:2012}. Discovery of the first planet, PDS 70 b, was announced in \citealt{Keppler:2018}, using data from VLT/SPHERE, VLT/NaCo, and Gemini/NICI. An additional planet, PDS 70 c, was identified the following year with MUSE \citep{Haffert:2019}. \citealt{Isella:2019} further identified continuum emission associated with PDS 70 c observations from ALMA, emission that they tentatively attributed to a circumplanetary disk. The CPD was not yet resolved from the surrounding protoplanetary disk, however.

\vspace{0.5em}

With additional, higher (20 millarcsecond) resolution observations from ALMA, \citealt{Benisty:2021} was finally able to resolve these features, showing the CPD emission as distinct from the protoplanetary disk. The planet is estimated to be about 2 Jupiter masses at a distance of $\sim$ 34 AU from the host star, with the CPD estimated to have a dust mass around 1-3\% $M_{\oplus}$ and a radial extent between 0.58 and 1.2 AU. PDS 70 c thus represents the first detection of a site of active moon formation (though the moons themselves have not yet been seen). It seems likely this system will continue to be an exciting target as a laboratory for testing moon-formation theories in the years to come, and may even be ripe for direct imaging of moons in the future.

\vspace{1.0em}

\noindent \textbf{2MASS J11193254–1137466 AB (2021).} \citealt{Limbach:2021} proposed that isolated planetary mass objects seen by their own light may be attractive targets to observe transiting satellites. These objects may be planets ejected from their natal stellar systems, in which case, it is thought they have $\gtrsim$ 50\% chance of retaining satellites, particularly those with small semi-major axes \citep[][]{Hong:2018, Rabago:2019}. An object with an orbit roughly equivalent to that of Io could have a geometric transit probability upwards of 10\%, and with short orbital periods, there could be numerous opportunities to observe transits.

\vspace{0.5em}

As a proof of concept, the team studied light curves from \textit{Spitzer} of two objects,  2MASS J1119-1137AB and WISEA J1147-2040, finding a ``fading event'' in the 2MASS J1119-1137AB light curve that they argued could be due to the transit of a 1.7 $R_{\oplus}$ companion. However, as the authors pointed out, the data show substantial variability with amplitudes comparable to that of the putative transit, making it difficult to rule out surface activity as the cause of the dimming event. Follow-up observations, to look for additional transits or to measure a radial velocity, could help disentangle these two possible scenarios. Multi-band photometry would be particularly worthwhile to help disentangle chromatic and achromatic events. In any case, the team argued that this potential detection demonstrates the feasibility of carrying out such observations, and that higher precision observations with JWST to search for these events have a decent shot at success.

\vspace{1.0em}

%\subsubsection*{Kepler-1708 b (2022)}
\noindent \textbf{Kepler-1708 b (2022).}
\citealt{k1708} announced the identification of a second transiting exomoon candidate, Kepler-1708 b-i. Like Kepler-1625 b-i, the candidate was identified as part of a broad survey of \textit{Kepler} systems, this time focusing on long period planets. The planet is substantially farther from its host star than Kepler-1625 b, however, with an orbital period of 737 days (compared to 287 days for Kepler-1625 b), and displays only two transits in the \textit{Kepler} data. But like Kepler-1625 b, Kepler-1708 b is also estimated to be a few Jupiter masses, and intriguingly, the moon candidate is also quite large, with a radius $R_S = 2.61^{+0.42}_{-0.43} \, R_{\oplus}$.

\vspace{0.5em}

This candidate detection raises some interesting questions: might large moons like 1625 b-i and 1708 b-i be more common than we had anticipated? Or are these, like Hot Jupiters before them, simply the lowest hanging fruit? Does the identification of a second large exomoon candidate lend additional credence to the first one being real? Or might they both be illusory, little more than red noise? \citealt{k1708} went to great lengths to investigate and rule out alternative explanations for the observed signals, but true confirmation may be years away. Recently, \citealt{Cassese:2022} found that Kepler-1708 b-i would likely be undetectable with HST, leaving JWST as the instrument of choice for a confirmation. As only two transits of the planet have been observed so far, additional transits will be important to observe, in particular, so that TTVs can be measured. These TTVs will provide an independent dynamical test for the moon hypothesis, can help to constrain a planet-moon mass ratio, and should improve future observations by refining the orbital period of the planet.

\vspace{1.0em}

\noindent \textbf{K2-33b (2022).} Recently, \citealt{Ohno:2022b} analyzed a transmission spectrum of the young (11 Myr) planet K2-33b produced from photometry sourced from \textit{K2}, MEarth, \textit{Spitzer}, and \textit{Hubble}. Observations suggested an exceptionally steep slope in the changing transit depths that was unsatisfactorily explained by other possible explanations, including star spots or aerosols. The authors found that a circumplanetary ring, one that is optically thick at shorter wavelengths and increasingly transparent at longer wavelengths, could fit the data well. On the other hand, \citealt{Thao:2023}, working concurrently, found that a tholin haze in the atmosphere could also result in the observed spectrum. Crucially, the ring model from \citealt{Ohno:2022b} predicts a strong silicate feature at 10 $\mu$m, which could be observed by JWST-MIRI. 

\vspace{1.0em}

%% file: characterization.tex
\label{sec:characterization}

\section{Characterization}
Once an exomoon or exoring candidate signal has been identified and various false positive scenarios have been ruled out (to the extent possible), our next imperative is to analyze the system architecture, first to determine stability and plausibility, then ultimately to better understand these objects and the Universe that gave rise to them.

\subsection{Asessing an Exomoon Candidate's Stability}
A viable exomoon candidate will have a model solution for its orbital parameters. We must then ask, are these solutions viable long-term? Could they have survived the age of the system? These investigations are not necessarily trivial, as they may involve very long integration times. The survivability of moons will be impacted by 1) the relationship of the moon to its planet, 2) the relationship of the moon to other moons in the system, 3) the relationship of the planet to other planets in the system, and 4) the relationship of the planet-moon system to the host star. Incorporating all these effects into a test of stability would be rather complex, so the researcher will have to decide which effects to include and which can be safely ignored for a given problem. Let us examine each of them in turn.

\vspace{0.5em}

To first approximation a moon will be stable if it is found somewhere between the Roche limit and a distance from its host planet that is some fraction of the Hill radius \citep{Domingos:2006, RosarioFranco:2020}. The latter is, of course, computed in relation to the host star, so we can never be rid of that consideration completely even if the star itself is left out of the simulation. In addition, the moon's inclination, and therefore its location within the planet's non-spherical gravitational potential, will also play a role, as will important physical parameters such as the Love number $k_2$ and the tidal quality factor $Q$ (a parameter that remains poorly constrained even for Solar System objects), and the evolving structure of the planet itself \citep{AlvaradoMontes:2017}. Taking these factors into account, though, introduces several additional unknowns that are generally not observable for the planet, such as its internal structure, its axial tilt, its rotation speed, and its oblateness. However, it may be that the identification of a moon in a given system could allow us to work backward to constrain some of these otherwise inaccessible values. 

\vspace{0.5em}

With respect to stability based on moon-moon interactions, we may desire to simulate these as isolated from stellar perturbations also, but this approximation will quite naturally become less reliable at smaller planet semi-major axes or significant planet eccentricities. Systems containing more than one moon can generally no longer be handled analytically, so an $N-$body integrator such as \textsc{Rebound} \citep{Rein:2012} will typically be required. The stability of the system may be assessed by running one or perhaps many simulations for a period of time long enough that long-term stability can be demonstrated (say, $10^9$ orbits), or stability metrics may be employed to assess the system's long-term viability \citep[e.g.][]{Cincotta:2003, Laskar:2017, Tamayo:2020}.

\vspace{0.5em}

Then there are planet-planet interactions. Neighboring planets will interact with one another gravitationally, even at great distances from one another; recall for example the discovery of Neptune, famously predicted due to an anomaly in the predicted positions of Uranus. Of course, planets at greater distances from one another will feel smaller perturbations, while planets at or near orbital resonance may experience more significant forces \citep[see][]{Deck:2013, Holczer:2016, Hadden:2017, Petit:2018}. Naturally, any force exerted on a planet by a neighbor will be felt by the moons as well, so once again the degree to which these forces may impact the moon system's stability will have to be assessed before any of these effects can be ignored.

\vspace{0.5em}

Mutual perturbations may be far more extreme in the early stages of planet formation, and may affect the fate of any moons. \citealt{Martinez:2019}, for example, examined the possibility of moons being liberated from their host planets via the Kozai-Lidov mechanism, as an external perturbing planet or brown dwarf excites the planet to high eccentricity, thereby producing a Hill region that is oscillating drastically. A significant fraction of these moons may survive to become planets in their own right, detectable in some cases with traces of their history still intact. \citealt{Gong:2013} examined the impact of planet-planet scattering events, finding (perhaps not surprisingly) that the events leading up to the creation of highly elliptical planets typically results in the destruction of any attendant moons, making planets orbiting with high eccentricities even less attractive targets in the moon search. \citealt{Hong:2018} likewise found that the moons might be destroyed during scattering events, but might also become planets in their own right, or even survive as the moon of a free-floating planet (see also \citealt{Rabago:2019}).

\vspace{0.5em}

Finally, we may consider scenarios that are impacted by the proximity of the planet to its host star. As planets migrate inward, the Hill spheres of the planets will shrink, with the potential to destabilize the moons \citep{Namouni:2010}. \citealt{Spalding:2016} further identified the possibility of moon loss through eccentricity growth via evection resonance encounters. However, particularly massive moons may in fact inhibit planet migration in some cases \citep{Trani:2020}.

\vspace{0.5em}

After migrating inward, the moons remain vulnerable to loss. A study from \citealt{Barnes:2002} investigated the survivability of moons around close-in planets, suggesting that massive moons could not survive around Hot Jupiters, but smaller (undetectable) moons may be longer-lived. Not only are Hill spheres quite small in this regime, but tidal interactions between the planet and moon will be such that an exchange of angular momentum causes the planet to spin up and the moon's orbit to decay. Massive moons raise stronger tides in the planet, thus increasing the angular momentum tranfer. On the other hand, \citealt{Cassidy:2009} argued that tidal decay may not be nearly so much of a problem, as values for $Q$ may have been underestimated by as much as eight orders of magnitude in \citealt{Barnes:2002}, making even Earth-sized moons stable in this extreme environment. \citealt{Sasaki:2012} agreed with the conclusion of \citealt{Barnes:2002}, but only in cases where no tidal locking was involved; by contrast, heavier moons would survive longer in cases where the planet is tidally locked to the moon, or to the star. That work provided an analytical framework to compute the lifetime of star-planet-moon systems, obviating the need for expensive $N$-body simulations in some important cases. Recent exomoon claims have been evaluated within this framework \citep{Quarles:2020, Tokadjian:2020}.  

\vspace{0.5em}

\citealt{Sucerquia:2019} employed the framework from \citealt{AlvaradoMontes:2017} to examine circumstances by which the moon's orbit is raised, not lowered, to the point where they may be liberated to become ``ploonets'', and suggested these objects could be detectable for tens of millions of years or more, possibly displaying signs of evaporation. \citealt{Dobos:2021} found in applying the \citealt{Barnes:2002} framework a strong correlation between the semi-major axis of the planet and the survival rate of moons, suggesting again that Hot Jupiters are not ideal targets for the exomoon search, with no moons surviving interior of $P=10$ days. \citealt{Makarov:2023} likewise found that among the currently known exoplanets, mostly short period, exomoons should be rare. These predictions are corroborated observationally by the low occurrence rate of exomoons found in \citealt{HEKVI}. Taken together, these studies suggest that a better understanding of $Q$ factors in particular may be key to determining the fate of close-in moons. And indeed, finding close-in moons could provide insights into this question, as well.

\vspace{0.5em}

In any case, Hot Jupiters are among the easiest exoplanets to study, owing to their large size and especially to their exceptionally short periods, offering myriad opportunities for observation. They also benefit from a large number of archived observations on premier telescopes. As such, even if they are \textit{a priori} unlikely to be moon hosts, the search for moons around these planets can be carried out at comparatively minor cost.

\subsection{Establishing a Plausible Exomoon Formation Pathway}
Assuming we have established a system's stability, the next question is, is there a plausible formation pathway? Indeed, this was the first question that arose from the claim of a Neptune-sized moon in the Kepler-1625 b system. Observers will naturally be inclined to let the observations speak for themselves; like Hot Jupiters before them, large moons are not impossible simply because we had not anticipated them. If the data say they are there, they are there, and the theorists have some work to do to explain them.

\vspace{0.5em}

The problem, of course, is that sometimes the data may be less than entirely clear. That is, we may be seeing \textit{hints} of an exomoon, but a firm confirmation remains somewhat elusive. Moreover, opportunities for confirmation may be rare; long period planets may be seen to transit only once every few years. Working at the frontiers of discovery, the lingering doubts and ambiguities are somewhat to be expected. Still, anything that is not readily explainable through known pathways will have a higher bar to clear for acceptance. In time, however, theorists may yet identify additional pathways to moon formation, as some have done already in the wake of the Kepler-1625 b findings, making initially exotic objects seem less so in time, and possibly even inspiring the search for objects that were once unthinkable.

\vspace{0.5em}

One key challenge is the sometimes large uncertainties on inferred system parameters. Planet masses may be poorly constrained, there may be a wide range of possible satellite inclinations, and orbital period solutions can be multi-modal, owing to the undersampling of the moon's true orbital frequency. In practice it may be quite challenging to simulate such a broad range of system properties efficiently or comprehensively. Theorists may therefore have to make some decisions as to what is most informative to investigate.

\vspace{0.5em}

To test a CPD formation scenario, theorists may opt to simulate disks of material (in isolation, or in relation to a protoplanetary disk), to examine how readily a moon or moon architecture can be made. The physics of these disks is complicated, involving gas and dust, turbulence, chemistry, electrodynamics, magnetic fields... as such, analytical handlings may be informative but ultimately insufficient; they may therefore be replaced by or complemented with $N-$body simulations 
\citep[e.g.][]{Ogihara:2012, Hyodo:2015, Fujii:2020, Batygin:2020, Cilibrasi:2021} or hydrodynamical simulations \citep[e.g.][]{Morbidelli:2007, Shabram:2013, Szulagyi:2017b}. Capture scenarios will likely utilize $N-$body simulations also, either simulating an entire planetary system \citep[e.g.][]{Nesvorny:2007}, or zooming in on the encounter of interest and examining the range of possible configurations that give rise to the capture \citep[e.g.][]{Hamers:2018, Hansen:2019}. Impact scenarios, likewise, may take either approach \citep[e.g.][]{Agnor:1999, Obrien:2014}. These simulations thus serve to show whether a moon in question has a plausible formation pathway, and ideally provide some sense of its probability.

\subsection{Addressing a Exomoon Candidate's Implications}
Having demonstrated the credibility of the signal, the plausibility of the moon hypothesis, and the long-term viability of the system, the last step is interpretation. What can the moon or moons tell us about the population? Where are moons most likely to be found? How large are they? How are they typically formed?  

\vspace{0.5em}

In time, as the number of detected moons (hopefully!) grows, a picture of the exomoon population and their occurrence rate as a function of planetary semi-major axis should begin to emerge. This will provide an important empirical test for determining the plausibility of future moon discoveries; a moon found where moons tend to be found will attract considerably less attention and skepticism than those that show up in the unpopulated regions of parameter space. In any case, these moons collectively should begin to provide additional insights into the migrational history of the planetary systems in which they are embedded.

\vspace{0.5em}

Confirmation of giant moon candidates like Kepler-1625 b-i and Kepler-1708 b-i could provide powerful evidence that moons need not necessarily resemble those in our Solar System. Estimates on the number of moons that may be found in the \textit{Kepler} data, based on (for example) the distribution of stellar spectral types, system distances, relative abundances of short- versus long-period planets, must always have a moon radius in mind. How many, say, Earth-radius moons might we see? Or Ganymede-radius moons? There will be a trade off here, as illustrated in \citealt{HEKVI}; if there is an abundance of large moons, we should be able to see them. Smaller moons will not be so easily seen, so there may be many of them but we just cannot see them.

\vspace{0.5em}

Identification of candidate signals for these giant moons raise an important question: if we are sensitive to such moons, what is their occurrence rate? Should we not have seen more of them by now? Or are they (as might be expected) quite rare? What does it mean that we have seen these putative moons, so far, only in star systems that are somewhat faint? Must we look out across vast distances in order to stumble upon these rare worlds? Or is the detection merely illusory, a product of that faintness, a mere ripple in the noise profile that would disappear with a cleaner data set? 

\vspace{0.5em}

These are not easy questions, but they will be raised, either by the researcher or their critics. And they are not always easy to answer. However unsettling, the fact is these questions may not be resolvable in the immediate term. We can look forward to new and less expensive observations in the future, but in the near term, we may simply have to resign ourselves to a situation where moon hypotheses may go unproved for several years. Situated somewhere between speculation and certainty, we will be left with these hints of what may be out there. But hopefully, with enough hints, a picture of the truth may begin to emerge.

\subsection{Assessing the Plausibility of an Exoring}

Claims that an observation is \textit{consistent} with a predicted phenomenon are insufficient for demonstrating that the phenomenon is the best explanation. Competing hypotheses need to be examined and ruled out to the extent possible, and importantly, the \textit{plausibility} of each scenario must be weighed. This is not always an easy task in the case of exoplanetary systems, because we continue to find systems that have no Solar System analogues and in some cases no precedent at all. 

\vspace{0.5em}

Determining the possibility of icy rings is first a matter of calculating the amount of incident flux received at a given distance from the host star and computing whether the ice under consideration (not necessarily water ice) can survive. If icy rings are impossible for a given semi-major axis, dusty or rocky rings may be considered, but it must be borne in mind their relevant observational properties going forward (for example, chromatic effects).

\vspace{0.5em}

Having determined the (im)possibility of certain ring compositions, one might move on to $N$-body simulations of the system, running the clock forward or backward in time to see whether observed properties can be either produced, or sustained, over long timescales. Again, because we generally do not expect to observe extremely short-lived events with much frequency, architectures that go unstable on short timescales must be treated with a good amount of skepticism. For example, a dynamical scenario that is found to be stable for only $10^5$ years existing in a system that has a 10 billion year lifetime would suggest we just happening to be seeing that system in the most interesting 0.001\% of its lifetime, which is highly improbable. In the case of rings, it may be that disk morphology can change substantially on a short timescale, as seen in \citealt{Sutton:2019}. Generally, high eccentricities for the planet will be reason for concern, especially if the planet passes close to its host star, as substantial tides on the system may be experienced leading to continual disturbances that may be destabilizing.

\vspace{0.5em}

After one has established the present-day plausibility of a ring system, formation pathways may be considered. But as there are a variety of mechanisms that can result in the production of a ring, demonstrating the presence of a ring to the community's satisfaction will probably rely less on demonstrating whether it is possible to form and more on whether 1) the data can support the ring hypothesis, and 2) other less exotic explanations can be excluded.

%% file: challenges.tex
\label{challenges}

\section{Ongoing Challenges}
It is worthwhile to highlight briefly some of the lingering challenges with respect to the exomoon search, in order that we might work collectively to find solutions to these obstacles. Let us now touch briefly on a few of these.

\subsection{Telescope Limitations}
We are enormously reliant on the nature and availability of telescope resources in the search for exomoons and exorings. For this work, we may turn to both ground- and space-based telescopes, tasked with either survey or target observations. Each of these will have their benefits and drawbacks. 

\vspace{0.5em}

The space-based \textit{Kepler} survey has been responsible for arguably the most promising candidates identified to date. For the exomoon search especially, we want to see as many transits of the planet-moon system as possible, and so the long time baseline of \textit{Kepler} (4.5 years in its primary mission) has enabled us to see many transits of short period planets and even a handful of planets beyond 1 AU. Unfortunately, there is a trade-off between time coverage and sky coverage; for the exoplanet search, \textit{Kepler}'s time coverage is unmatched, but it monitored a small patch of the sky, so we are limited to a few thousand transiting planets identified in a sample of a couple hundred thousand stars. This sample is certainly large, but when we begin layering on the additional requirements to see exomoons -- we want transiting at large semi-major axes (a small percentage of), with comparatively bright, but not large, stars (a small percentage of those) -- we run out of attractive targets rather quickly.

\vspace{0.5em}

TESS, on the other hand, covers nearly the entire sky, but the vast majority of its coverage lasts only 27.5 days. A fraction of the sky at higher ecliptic latitudes gets longer baselines, however, owing to overlap of the detectors' footprint on the sky, reaching up to a full year at the ecliptic poles (the Continuous Viewing Zone). Thus, despite the enormous number of stars for which we have data, those long-period planets we are particularly interested in are even more rare to find.

\vspace{0.5em}

For objects that are unchanging on the time scale of the observation, one needs only to increase the exposure time to improve the signal-to-noise ratio. But this simply does not work for observing events in the time-domain; longer exposures will shrink the uncertainties in flux, but come at the expense of time resolution, and ultimately do not improve our sensitivity \citep[e.g.][]{Kipping:2010}. Therefore, in cases where uncertainties are shot noise limited, only a larger aperture can reduce uncertainties and help us see shallower transits. This is what makes instruments such as HST and JWST particularly important for the search. 

\vspace{0.5em}

Ground-based telescopes clearly do not have the size and weight limitations of spacecraft, and may therefore have much larger mirrors and are capable of achieving superior photometric precision under ideal conditions and atmospheric interference mitigation. Unfortunately, there is trade-off: the rotation of the Earth makes long-duration observations impossible. Long-period planets, which appear to represent the most attractive targets for the exomoon search, have correspondingly long transits which are inappropriate for ground-based telescopes. 

\vspace{0.5em}

Moreover, we will generally want to observe the star for a long time before and after the planet's transit, not only to search for an exomoon transit signal, but also to calibrate the brightness of the star and to get a handle on any astrophysical variation present. Both observation-long trends and short-duration variation may be present, and these can have a major impact on our handling of the data. Consider the case of Kepler-1625 b: a full 40 hours of uninterrupted observations of the star were performed with HST, capturing roughly 19 hours of in-transit data and 21 hours of out-of-transit data (10-11 hours on either side of the planet's transit). Yet with this unprecedented observation, significant ambiguities remain, due in part to the challenges of handling astrophysical systematics in a finite time window. The problem is only exacerbated for ground-based observations, where we can only hope to monitor an event for a mere fraction of the time that was allotted on HST for the Kepler-1625 b transit. 

\vspace{0.5em}

One mitigating strategy to improve time coverage might be to utilize telescopes all around the world in a coordinated observational effort. This clearly comes with its own challenges: we require participation from observers all around the world, favorable weather conditions in multiple locations, and possibly, multiple observation proposals to all be awarded. This is a tall order, but one exciting possibility is that citizen scientists may be enlisted to join the effort, as was recently carried out to observe a transit of Kepler-167 e with Unistellar eVscopes \citep{unistellar_k167}. Combining these datasets will come with considerable challenges, but in time we may be able to overcome them to produce datasets that can now only be achievable with targeted, space-based observations. We can of course only expect so much from small-aperture, consumer-grade equipment, but uninterrupted observations from the ground should still be of significant value, especially for the maintenance of ephemerides or the measurement of TTVs.

\vspace{0.5em}

Whether we are observing these transits from the ground or in space, obtaining dedicated observations will remain a significant obstacle. Convincing a time allocation committee that these observations have a good chance of succeeding, and that the objective is worthwhile, is no easy feat, especially when there is not yet significant buy-in from the community. What's more, the minimum time-coverage needed for observing a single exomoon transit event may be considerably longer than that which is typically required for lower-risk science, making it even more difficult for a TAC to justify. Nevertheless, we will have to continue these efforts if we hope to achieve better precision than that which is available on survey telescopes, and to the extent possible, we will have to identify ways of mitigating risk and acquiring critical data in such a way that conserves resources.

\subsection{Modeling Challenges}

\subsubsection{Inclined Moon Problem}
Moons in our Solar System are found with a wide range of orbital inclinations, some orbits carrying the moons far away from the planet's orbital plane. We can reasonably assume exomoons may have similar architectures. Depending on the orientation and semi-major axis of the exomoon, then, it is quite possible for exomoons not to transit every time the planet does; in such a geometry the moon may pass above or below the face of the star and not block out any light. \citealt{Martin:2019} presented a helpful mathematical framework for computing transit probabilities based on this geometry and highlighted its impact on observational strategies. They calculated, for example, that based on published values for Kepler-1625 b-i, the moon may only transit 40\% of the time. As such, one would need to observe six transits of the planet in a row, all without evidence of a moon transit, to refute the presence of the moon to 95\% confidence. Needless to say, that is a lot of time for a targeted observation campaign, and unfortunately, we are unlikely to see any survey observations of this field with sufficient sensitivity for a long time to come.

\vspace{0.5em}

One particularly troubling aspect of these inclined moons is that we have little reason to reject an inclined moon solution. Particularly if we are considering captured moons, which large moons may very well be, large inclinations are entirely reasonable. But such a situation requires only a fraction of planetary transits to display evidence of a moon. It is conceivable that only a fraction of transits displaying a moon-like flux dip could be sufficient to claim evidence of a moon, and yet this is not entirely satisfactory. It is almost too convenient. Indeed, this was the case for the transits of Kepler-1625 b, with the results favoring a moon but strongly depending on the presence of a moon-like dip in the HST data by itself.

\vspace{0.5em}

Might we be able to do better? In addition to testing a moon model with inclination as a free parameter, we can certainly test an additional model that forces co-planarity of the moon's orbit with that of the planet, but this is contrived. We might do a bit better by having a prior on moon inclinations, such that edge-on planets are simply \textit{preferred} over high-inclination planets. But where do we get this prior? In the near-term we can hardly do better than testing a variety of priors and examining the results, but a theoretical underpinning would be more satisfying, and ultimately, we would want these priors to have more of an empirical basis. Unfortunately, the Solar System seems too small of a sample size be very informative, and in light of the possibility that exomoons can look very different from Solar System moons, we would need to be careful about extrapolating the Solar System sample anyway.

\subsubsection{Multiple Moons and Priors}
Another key observation is that moons are rarely alone; more often a planet hosts several or even many moons. Despite this, fitting an exomoon model to real data has virtually always been done with a single moon. There have been good reasons for this: it is a simpler model and more computationally expendient. Each additional moon in the model also requires seven additional free parameters, and this also presents us with something of a dilemma. 

\vspace{0.5em}

It is well known that a model with more free parameters can more readily fit a data set. Thus, a model with more moons will be virtually guaranteed to fit the observations better than a model with fewer moons, and therefore must be penalized. This is mathematically and statistically sound, but a system with multiple moons is not more complex in terms of being a more contrived model. It may even be more likely \textit{a priori}. In penalizing these multi-moon models, we may be biasing ourselves against systems that are in fact hosting multiple moons. How might we overcome this? As always, more informative priors can help. Again, a theoretical underpinning will be worthwhile, and in time we may hope to be guided by real observations of the exomoon population.

\subsubsection{Single-Transiting Planets}
Planets orbiting at large distances from their host star are among the most attractive targets, from a physical standpoint, to search for exomoons. As previously mentioned, this expectation is based both on analogy to the Solar System, and with an emerging picture of the exomoon population \citep[e.g.][]{Spalding:2016, HEKVI, Sucerquia:2020, Dobos:2021}. But if this is true, we do not only need to contend with long and infrequent transit events; for some planets, we may have observed only a single transit of the planet. As such, there is significant uncertainty in the orbital period of the planet \citep{Osborn:2016, Sandford:2019}. We must do better than this if we have any hope of observing these planets a second time. 

\vspace{0.5em}

The bluntest instrument, perhaps, would be to design an observing campaign that continuously checks for transits of these events, something akin to the MEarth project \citep{MEarth} for monitoring M-dwarfs in search of transiting planets. With enough long-period planets in the catalogue, we can imagine an automated system that would regularly check in on these planets and kick into high gear if a transit is detected. Once a second transit is detected, the orbital period is now essentially known, and follow-up observations with more sensitive telescopes can be planned for the future.

\vspace{0.5em}

But perhaps we can do better. The morphology of a planet's transit encodes several critical pieces of information \citep[e.g.][]{MandelAgol:2002}: the transit duration tells us how long it takes for the planet to pass in front of the face of the star; the star's mass and radius can be inferred from stellar models, or the density can be computed from the light curve itself \citep{Sandford:2019}; the planet's radius may be calculated from the transit depth, and the mass from a mass-radius relationship \citep[e.g.][]{forecaster}; the impact parameter may be inferred if we have good enough constraints on the limb darkening of the star, which will affect the shapes of planetary ingress and egress; We may further infer the planet's eccentricity, and its true anomaly, recognizing that the planet will have different orbital velocity during ingress and egress and this too will imprint on the shape of the transit \citep{photoeccentric}. Of course, these are not trivial to recover, will each come with their own uncertainties, and will be dependent on the quality of the photometry. Nevertheless, we may hope for some innovations along these lines, perhaps by incorporating a variety of other observables, machine learning, and a significantly larger data set, such that single-transiting planets can have their orbital periods predicted with significantly greater accuracy in the years to come. Targeted follow-up observations will require highly accurate ephemerides, so this is a major priority.

%% file: conclusion.tex
\label{sec:conclusion}

\section{Conclusion}
We have now come to the end of this review, and hopefully the reader has some new appreciation for some of the exciting opportunities in exomoon and exoring science as well as some of the challenges that we as a community will need to overcome. We began with survey of our expectations, and moved to discussing some observational signatures that we might go look for. We then turned to examine previous efforts to look for these signatures, and to highlight a few important case studies in the effort. From there we touched on how we might go about validating and characterizing these exomoons, and explored some complicating factors in our search.

\vspace{0.5em}

The good news is, there is no shortage of work to be done. That's good news to a scientist, anyway. The bad news is, we may still have a long way to go before exomoon and exoring discoveries will be as routine as exoplanet discoveries are now, and there is little telling what that timeline might look like. What's needed is a combination of future surveys, dedicated observations, researchers willing to invest the effort, and continued innovation. It is not easy to predict where imaginations may take us; if someone already knew how to improve our chances of finding these objects, we would already be trying it. Yet, as we survey the history of science, we are reminded that ingenuity knows no bounds. Hardly anyone doubts there are moons and rings to be found out there; it's just a matter of time before we find them. What is clear is that the eventual payoff for finding and characterizing these worlds will be substantial.

%% file: xref.tex
\label{sec:conclusion}

\section{Cross-References}
\begin{itemize}
    \item Rings in the Solar System: A Short Review

    \item Transit Photometry as an Exoplanet Discovery Method

    \item Transit-Timing and Duration Variations for the Discovery and Characterization of Exoplanets

    \item Space Missions for Extrasolar Planets: Overview and Introduction

    \item The Impact of Stellar Activity on the Detection and Characterization of Exoplanets

    \item A Brief Overview of Planet Formation

    \item Formation of Giant Planets

    \item Dynamical Evolution of Planetary Systems

    \item The Habitability of Icy Ocean Worlds in the Solar System

    \item Special Cases: Moons, Rings, Comets, and Trojans 
\end{itemize}

%% file: bibliography.tex
\begin{multicols}{2}
\def\bibfont{\footnotesize}

\end{multicols}

%% file: main.bbl
\begin{thebibliography}{}

%%% ALPHABETICAL VERSION
%\begin{comment}

\bibitem[Agnor et al.(1999)]{Agnor:1999} Agnor, C.~B., Canup, R.~M., \& Levison, H.~F.\ 1999, \icarus, 142, 219. doi:10.1006/icar.1999.6201

\vspace{-1em}

\bibitem[Agnor \& Hamilton(2006)]{Agnor:2006} Agnor, C.~B. \& Hamilton, D.~P.\ 2006, \nat, 441, 192. doi:10.1038/nature04792

\vspace{-1em}

\bibitem[Agol et al.(2015)]{Agol:2015} Agol, E., Jansen, T., Lacy, B., et al.\ 2015, \apj, 812, 5. doi:10.1088/0004-637X/812/1/5

\vspace{-1em}

\bibitem[Aizawa et al.(2017)]{Aizawa:2017} Aizawa, M., Uehara, S., Masuda, K., et al.\ 2017, \aj, 153, 193. doi:10.3847/1538-3881/aa6336

\vspace{-1em}

\bibitem[Alshehhi et al.(2020)]{Alshehhi:2020} Alshehhi, R., Rodenbeck, K., Gizon, L., et al.\ 2020, \aap, 640, A41. doi:10.1051/0004-6361/201937059

\vspace{-1em}

\bibitem[Alvarado-Montes et al.(2017)]{AlvaradoMontes:2017} Alvarado-Montes, J.~A., Zuluaga, J.~I., \& Sucerquia, M.\ 2017, \mnras, 471, 3019. doi:10.1093/mnras/stx1745

\vspace{-1em}

\bibitem[Arnold \& Schneider(2004)]{Arnold:2004} Arnold, L. \& Schneider, J.\ 2004, \aap, 420, 1153. doi:10.1051/0004-6361:20035720

\vspace{-1em}

\bibitem[{\'A}vila et al.(2021)]{Avila:2021} {\'A}vila, P.~J., Grassi, T., Bovino, S., et al.\ 2021, International Journal of Astrobiology, 20, 300. doi:10.1017/S1473550421000173

\vspace{-1em}

\bibitem[Bachelet et al.(2022)]{Bachelet:2022} Bachelet, E., Specht, D., Penny, M., et al.\ 2022, \aap, 664, A136. doi:10.1051/0004-6361/202140351

\vspace{-1em}

\bibitem[Barclay et al.(2013)]{Barclay:2013} Barclay, T., Rowe, J.~F., Lissauer, J.~J., et al.\ 2013, \nat, 494, 452. doi:10.1038/nature11914

\vspace{-1em}

\bibitem[Barnes \& O'Brien(2002)]{Barnes:2002} Barnes, J.~W. \& O'Brien, D.~P.\ 2002, \apj, 575, 1087. doi:10.1086/341477

\vspace{-1em}

\bibitem[Barnes \& Fortney(2004)]{Barnes:2004} Barnes, J.~W. \& Fortney, J.~J.\ 2004, \apj, 616, 1193. doi:10.1086/425067

\vspace{-1em}

\bibitem[Batygin \& Morbidelli(2020)]{Batygin:2020} Batygin, K. \& Morbidelli, A.\ 2020, \apj, 894, 143. doi:10.3847/1538-4357/ab8937

\vspace{-1em}

\bibitem[Benisty et al.(2021)]{Benisty:2021} Benisty, M., Bae, J., Facchini, S., et al.\ 2021, \apjl, 916, L2. doi:10.3847/2041-8213/ac0f83

\vspace{-1em}

\bibitem[Ben-Jaffel \& Ballester(2014)]{BenJaffel:2014} Ben-Jaffel, L. \& Ballester, G.~E.\ 2014, \apjl, 785, L30. doi:10.1088/2041-8205/785/2/L30

\vspace{-1em}

\bibitem[Bennett et al.(2014)]{Bennett:2014} Bennett, D.~P., Batista, V., Bond, I.~A., et al.\ 2014, \apj, 785, 155. doi:10.1088/0004-637X/785/2/155

\vspace{-1em}

\bibitem[Bernstein et al.(2004)]{kuiper_belt_mass1} Bernstein, G.~M., Trilling, D.~E., Allen, R.~L., et al.\ 2004, \aj, 128, 1364. doi:10.1086/422919

\vspace{-1em}

\bibitem[Berta et al.(2012)]{MEarth} Berta, Z.~K., Irwin, J., Charbonneau, D., et al.\ 2012, \aj, 144, 145. doi:10.1088/0004-6256/144/5/145

\vspace{-1em}

\bibitem[Bierson et al.(2023)]{Io_dry} Bierson, C.~J., Fortney, J.~J., Trinh, K.~T., et al.\ 2023, \psj, 4, 122. doi:10.3847/PSJ/ace2c7

\vspace{-1em}

\bibitem[Borucki et al.(2011)]{Borucki:2011} Borucki, W.~J., Koch, D.~G., Basri, G., et al.\ 2011, \apj, 736, 19. doi:10.1088/0004-637X/736/1/19

\vspace{-1em}

\bibitem[Bottke et al.(2005)]{Jupiter_asteroids} Bottke, W.~F., Durda, D.~D., Nesvorn{\'y}, D., et al.\ 2005, \icarus, 175, 111. doi:10.1016/j.icarus.2004.10.026

\vspace{-1em}

\bibitem[Braga-Ribas et al.(2014)]{Chariklo_rings} Braga-Ribas, F., Sicardy, B., Ortiz, J.~L., et al.\ 2014, \nat, 508, 72. doi:10.1038/nature13155

\vspace{-1em}

\bibitem[Cabrera \& Schneider(2007)]{Cabrera:2007} Cabrera, J. \& Schneider, J.\ 2007, \aap, 464, 1133. doi:10.1051/0004-6361:20066111

\vspace{-1em}

\bibitem[Campante et al.(2015)]{Campante:2015} Campante, T.~L., Barclay, T., Swift, J.~J., et al.\ 2015, \apj, 799, 170. doi:10.1088/0004-637X/799/2/170

\vspace{-1em}

\bibitem[Canup \& Asphaug(2001)]{Canup:2001} Canup, R.~M., \& Asphaug, E.\ 2001, \nat, 412, 708

\vspace{-1em}

\bibitem[Canup \& Ward(2002)]{Canup:2002} Canup, R.~M. \& Ward, W.~R.\ 2002, \aj, 124, 3404. doi:10.1086/344684

\vspace{-1em}

\bibitem[Canup(2004)]{Canup:2004} Canup, R.~M.\ 2004, \icarus, 168, 433. doi:10.1016/j.icarus.2003.09.028

\vspace{-1em}

\bibitem[Canup \& Ward(2006)]{Canup:2006} Canup, R.~M., \& Ward, W.~R.\ 2006, \nat, 441, 834

\vspace{-1em}

\bibitem[Canup(2010)]{Saturn_moon_stripping} Canup, R.~M.\ 2010, \nat, 468, 943. doi:10.1038/nature09661

\vspace{-1em}

\bibitem[Canup(2011)]{Canup:2011} Canup, R.~M.\ 2011, \aj, 141, 35. doi:10.1088/0004-6256/141/2/35

\vspace{-1em}

\bibitem[Canup(2012)]{Canup:2012} Canup, R.~M.\ 2012, Science, 338, 1052. doi:10.1126/science.1226073

\vspace{-1em}

\bibitem[Carr et al.(1998)]{Carr:1998} Carr, M.~H., Belton, M.~J.~S., Chapman, C.~R., et al.\ 1998, \nat, 391, 363. doi:10.1038/34857

\vspace{-1em}

\bibitem[Carter et al.(2008)]{Carter:2008} Carter, J.~A., Yee, J.~C., Eastman, J., et al.\ 2008, \apj, 689, 499. doi:10.1086/592321

\vspace{-1em}

\bibitem[Carter \& Winn(2010)]{Carter:2010} Carter, J.~A. \& Winn, J.~N.\ 2010, \apj, 716, 850. doi:10.1088/0004-637X/716/1/850

\vspace{-1em}

\bibitem[Cassese \& Kipping(2022)]{Cassese:2022} Cassese, B. \& Kipping, D.\ 2022, \mnras, 516, 3701. doi:10.1093/mnras/stac2090

\vspace{-1em}

\bibitem[Cassidy et al.(2009)]{Cassidy:2009} Cassidy, T.~A., Mendez, R., Arras, P., et al.\ 2009, \apj, 704, 1341. doi:10.1088/0004-637X/704/2/1341

\vspace{-1em}

\bibitem[Chakraborty \& Kipping(2023)]{Chakraborty:2023} Chakraborty, J. \& Kipping, D.\ 2023, \mnras, 519, 2690. doi:10.1093/mnras/stac3604

\vspace{-1em}

\bibitem[Chauvin et al.(2004)]{Chauvin:2004} Chauvin, G., Lagrange, A.-M., Dumas, C., et al.\ 2004, \aap, 425, L29. doi:10.1051/0004-6361:200400056

\vspace{-1em}

\bibitem[Chen \& Kipping(2017)]{forecaster} Chen, J. \& Kipping, D.\ 2017, \apj, 834, 17. doi:10.3847/1538-4357/834/1/17

\vspace{-1em}

\bibitem[Chrenko et al.(2018)]{Chrenko:2018} Chrenko, O., Bro{\v{z}}, M., \& Nesvorn{\'y}, D.\ 2018, \apj, 868, 145

\vspace{-1em}

\bibitem[Cilibrasi et al.(2018)]{Cilibrasi:2018} Cilibrasi, M., Szul{\'a}gyi, J., Mayer, L., et al.\ 2018, \mnras, 480, 4355. doi:10.1093/mnras/sty2163

\vspace{-1em}

\bibitem[Cilibrasi et al.(2021)]{Cilibrasi:2021} Cilibrasi, M., Szul{\'a}gyi, J., Grimm, S.~L., et al.\ 2021, \mnras, 504, 5455. doi:10.1093/mnras/stab1179

\vspace{-1em}

\bibitem[Cincotta et al.(2003)]{Cincotta:2003} Cincotta, P.~M., Giordano, C.~M., \& Sim{\'o}, C.\ 2003, Physica D Nonlinear Phenomena, 182, 151. doi:10.1016/S0167-2789(03)00103-9

\vspace{-1em}

\bibitem[Dawson \& Johnson(2012)]{photoeccentric} Dawson, R.~I. \& Johnson, J.~A.\ 2012, \apj, 756, 122. doi:10.1088/0004-637X/756/2/122

\vspace{-1em}

\bibitem[Deck et al.(2013)]{Deck:2013} Deck, K.~M., Payne, M., \& Holman, M.~J.\ 2013, \apj, 774, 129. doi:10.1088/0004-637X/774/2/129

\vspace{-1em}

\bibitem[de Mooij et al.(2017)]{deMooij:2017} de Mooij, E.~J.~W., Watson, C.~A., \& Kenworthy, M.~A.\ 2017, \mnras, 472, 2713. doi:10.1093/mnras/stx2142

\vspace{-1em}

\bibitem[Dobos \& Turner(2015)]{Dobos:2015} Dobos, V. \& Turner, E.~L.\ 2015, \apj, 804, 41. doi:10.1088/0004-637X/804/1/41

\vspace{-1em}

\bibitem[Dobos et al.(2017)]{Dobos:2017} Dobos, V., Heller, R., \& Turner, E.~L.\ 2017, \aap, 601, A91. doi:10.1051/0004-6361/201730541

\vspace{-1em}

\bibitem[Dobos et al.(2021)]{Dobos:2021} Dobos, V., Charnoz, S., P{\'a}l, A., et al.\ 2021, \pasp, 133, 094401. doi:10.1088/1538-3873/abfe04

\vspace{-1em}

\bibitem[Domingos et al.(2006)]{Domingos:2006} Domingos, R.~C., Winter, O.~C., \& Yokoyama, T.\ 2006, \mnras, 373, 1227. doi:10.1111/j.1365-2966.2006.11104.x

\vspace{-1em}

\bibitem[Dones(1991)]{Saturn_centaur_stripping} Dones, L.\ 1991, \icarus, 92, 194. doi:10.1016/0019-1035(91)90045-U

\vspace{-1em}

\bibitem[Dragomir et al.(2020)]{Dragomir:2020} Dragomir, D., Harris, M., Pepper, J., et al.\ 2020, \aj, 159, 219. doi:10.3847/1538-3881/ab845d

\vspace{-1em}

\bibitem[Duncan \& Levison(1997)]{scattered_disk} Duncan, M.~J. \& Levison, H.~F.\ 1997, Science, 276, 1670. doi:10.1126/science.276.5319.1670

\vspace{-1em}

\bibitem[Elliot et al.(1977)]{Uranus_rings} Elliot, J.~L., Dunham, E., \& Mink, D.\ 1977, \nat, 267, 328. doi:10.1038/267328a0

\vspace{-1em}

\bibitem[Fabrycky \& Winn(2009)]{Fabrycky:2009} Fabrycky, D.~C. \& Winn, J.~N.\ 2009, \apj, 696, 1230. doi:10.1088/0004-637X/696/2/1230

\vspace{-1em}

\bibitem[Forgan \& Yotov(2014)]{Forgan:2014} Forgan, D. \& Yotov, V.\ 2014, \mnras, 441, 3513. doi:10.1093/mnras/stu740

\vspace{-1em}

\bibitem[Forgan \& Dobos(2016)]{Forgan:2016} Forgan, D. \& Dobos, V.\ 2016, \mnras, 457, 1233. doi:10.1093/mnras/stw024

\vspace{-1em}

\bibitem[Fox \& Wiegert(2021)]{Fox:2021} Fox, C. \& Wiegert, P.\ 2021, \mnras, 501, 2378. doi:10.1093/mnras/staa3743

\vspace{-1em}

\bibitem[Fujii \& Ogihara(2020)]{Fujii:2020} Fujii, Y.~I. \& Ogihara, M.\ 2020, \aap, 635, L4. doi:10.1051/0004-6361/201937192

\vspace{-1em}

\bibitem[Gao \& Zhang(2020)]{Gao:2020} Gao, P. \& Zhang, X.\ 2020, \apj, 890, 93. doi:10.3847/1538-4357/ab6a9b

\vspace{-1em}

\bibitem[Gebek \& Oza(2020)]{Gebek:2020} Gebek, A. \& Oza, A.~V.\ 2020, \mnras, 497, 5271. doi:10.1093/mnras/staa2193

\vspace{-1em}

\bibitem[Gillon et al.(2017)]{TRAPPIST1} Gillon, M., Triaud, A.~H.~M.~J., Demory, B.-O., et al.\ 2017, \nat, 542, 456. doi:10.1038/nature21360

\vspace{-1em}

\bibitem[Goldreich \& Tremaine(1982)]{Goldreich:1982} Goldreich, P. \& Tremaine, S.\ 1982, \araa, 20, 249. doi:10.1146/annurev.aa.20.090182.001341

\vspace{-1em}

\bibitem[Gomes et al.(2005)]{late_heavy_bombardment} Gomes, R., Levison, H.~F., Tsiganis, K., et al.\ 2005, \nat, 435, 466. doi:10.1038/nature03676

\vspace{-1em}

\bibitem[Gong et al.(2013)]{Gong:2013} Gong, Y.-X., Zhou, J.-L., Xie, J.-W., et al.\ 2013, \apjl, 769, L14. doi:10.1088/2041-8205/769/1/L14

\vspace{-1em}

\bibitem[Gordon et al.(2020)]{Gordon:2020} Gordon, T.~A., Agol, E., \& Foreman-Mackey, D.\ 2020, \aj, 160, 240. doi:10.3847/1538-3881/abbc16

\vspace{-1em}

\bibitem[Gordon \& Agol(2022)]{Gefera} Gordon, T.~A. \& Agol, E.\ 2022, \aj, 164, 111. doi:10.3847/1538-3881/ac82b1

\vspace{-1em}

\bibitem[Green et al.(2021)]{Green:2021} Green, J., Boardsen, S., \& Dong, C.\ 2021, \apjl, 907, L45. doi:10.3847/2041-8213/abd93a

\vspace{-1em}

\bibitem[Gregorio-Hetem et al.(1992)]{PDS70} Gregorio-Hetem, J., Lepine, J.~R.~D., Quast, G.~R., et al.\ 1992, \aj, 103, 549. doi:10.1086/116082

\vspace{-1em}

\bibitem[Hadden \& Lithwick(2017)]{Hadden:2017} Hadden, S. \& Lithwick, Y.\ 2017, \aj, 154, 5. doi:10.3847/1538-3881/aa71ef

\vspace{-1em}

\bibitem[Haffert et al.(2019)]{Haffert:2019} Haffert, S.~Y., Bohn, A.~J., de Boer, J., et al.\ 2019, Nature Astronomy, 3, 749. doi:10.1038/s41550-019-0780-5

\vspace{-1em}

\bibitem[Hamers \& Portegies Zwart(2018)]{Hamers:2018} Hamers, A.~S., \& Portegies Zwart, S.~F.\ 2018, \apjl, 869, L27

\vspace{-1em}

\bibitem[Hamers et al.(2018)]{Hamers:2018b} Hamers, A.~S., Cai, M.~X., Roa, J., et al.\ 2018, \mnras, 480, 3800. doi:10.1093/mnras/sty2117

\vspace{-1em}

\bibitem[Hansen et al.(2006)]{Hansen:2006} Hansen, C.~J., Esposito, L., Stewart, A.~I.~F., et al.\ 2006, Science, 311, 1422. doi:10.1126/science.1121254

\vspace{-1em}

\bibitem[Hansen(2019)]{Hansen:2019} Hansen, B.~M.~S.\ 2019, Science Advances, 5, eaaw8665. doi:10.1126/sciadv.aaw8665

\vspace{-1em}

\bibitem[Haqq-Misra \& Heller(2018)]{Haqq-Misra:2018} Haqq-Misra, J. \& Heller, R.\ 2018, \mnras, 479, 3477. doi:10.1093/mnras/sty1630

\vspace{-1em}

\bibitem[Hartmann \& Davis(1975)]{Hartmann:1975} Hartmann, W.~K. \& Davis, D.~R.\ 1975, \icarus, 24, 504. doi:10.1016/0019-1035(75)90070-6

\vspace{-1em}

\bibitem[Hashimoto et al.(2012)]{Hashimoto:2012} Hashimoto, J., Dong, R., Kudo, T., et al.\ 2012, \apjl, 758, L19. doi:10.1088/2041-8205/758/1/L19

\vspace{-1em}

\bibitem[Hatchett et al.(2018)]{Hatchett:2018} Hatchett, W.~T., Barnes, J.~W., Ahlers, J.~P., et al.\ 2018, \na, 60, 88. doi:10.1016/j.newast.2017.11.001

\vspace{-1em}

\bibitem[Hedman et al.(2009)]{Hedman:2009} Hedman, M.~M., Murray, C.~D., Cooper, N.~J., et al.\ 2009, \icarus, 199, 378. doi:10.1016/j.icarus.2008.11.001

\vspace{-1em}

\bibitem[Heising et al.(2015)]{Heising:2015} Heising, M.~Z., Marcy, G.~W., \& Schlichting, H.~E.\ 2015, \apj, 814, 81. doi:10.1088/0004-637X/814/1/81

\vspace{-1em}

\bibitem[Heller(2012)]{Heller:2012} Heller, R.\ 2012, \aap, 545, L8. doi:10.1051/0004-6361/201220003

\vspace{-1em}

\bibitem[Heller \& Zuluaga(2013)]{Heller:2013} Heller, R. \& Zuluaga, J.~I.\ 2013, \apjl, 776, L33. doi:10.1088/2041-8205/776/2/L33

\vspace{-1em}

\bibitem[Heller \& Barnes(2013)]{Heller:2013b} Heller, R. \& Barnes, R.\ 2013, Astrobiology, 13, 18. doi:10.1089/ast.2012.0859

\vspace{-1em}

\bibitem[Heller(2014)]{Heller:2014} Heller, R.\ 2014, \apj, 787, 14

\vspace{-1em}

\bibitem[Heller \& Albrecht(2014)]{Heller:2014b} Heller, R. \& Albrecht, S.\ 2014, \apjl, 796, L1. doi:10.1088/2041-8205/796/1/L1

\vspace{-1em}

\bibitem[Heller \& Pudritz(2015)]{Heller:2015} Heller, R. \& Pudritz, R.\ 2015, \apj, 806, 181. doi:10.1088/0004-637X/806/2/181

\vspace{-1em}

\bibitem[Heller et al.(2015)]{Heller:2015b} Heller, R., Marleau, G.-D., \& Pudritz, R.~E.\ 2015, \aap, 579, L4. doi:10.1051/0004-6361/201526348

\vspace{-1em}

\bibitem[Heller \& Barnes(2015)]{Heller:2015c} Heller, R. \& Barnes, R.\ 2015, International Journal of Astrobiology, 14, 335. doi:10.1017/S1473550413000463

\vspace{-1em}

\bibitem[Heller et al.(2016)]{Heller:2016} Heller, R., Hippke, M., \& Jackson, B.\ 2016, \apj, 820, 88. doi:10.3847/0004-637X/820/2/88

\vspace{-1em}

\bibitem[Heller et al.(2016)]{Heller:2016b} Heller, R., Hippke, M., Placek, B., et al.\ 2016, \aap, 591, A67. doi:10.1051/0004-6361/201628573

\vspace{-1em}

\bibitem[Heller(2018)]{Heller_K1625_formation} Heller, R.\ 2018, \aap, 610, A39. doi:10.1051/0004-6361/201731760

\vspace{-1em}

\bibitem[Heller et al.(2019)]{Heller:2019} Heller, R., Rodenbeck, K., \& Bruno, G.\ 2019, \aap, 624, A95. doi:10.1051/0004-6361/201834913

\vspace{-1em}

\bibitem[Heller \& Hippke(2022)]{Heller:2022} Heller, R. \& Hippke, M.\ 2022, \aap, 657, A119. doi:10.1051/0004-6361/202142403

\vspace{-1em}

\bibitem[Herrero et al.(2016)]{Herrero:2016} Herrero, E., Ribas, I., Jordi, C., et al.\ 2016, \aap, 586, A131. doi:10.1051/0004-6361/201425369

\vspace{-1em}

\bibitem[Hinkel \& Kane(2013)]{Hinkel:2013} Hinkel, N.~R. \& Kane, S.~R.\ 2013, \apj, 774, 27. doi:10.1088/0004-637X/774/1/27

\vspace{-1em}

\bibitem[Hippke(2015)]{Hippke:2015} Hippke, M.\ 2015, \apj, 806, 51. doi:10.1088/0004-637X/806/1/51

\vspace{-1em}

\bibitem[Hippke \& Heller(2022)]{Pandora} Hippke, M. \& Heller, R.\ 2022, \aap, 662, A37. doi:10.1051/0004-6361/202243129

\vspace{-1em}

\bibitem[Holczer et al.(2016)]{Holczer:2016} Holczer, T., Mazeh, T., Nachmani, G., et al.\ 2016, \apjs, 225, 9. doi:10.3847/0067-0049/225/1/9

\vspace{-1em}

\bibitem[Hong et al.(2018)]{Hong:2018} Hong, Y.-C., Raymond, S.~N., Nicholson, P.~D., et al.\ 2018, \apj, 852, 85. doi:10.3847/1538-4357/aaa0db

\vspace{-1em}

\bibitem[H{\"o}rst(2017)]{Horst:2017} H{\"o}rst, S.~M.\ 2017, Journal of Geophysical Research (Planets), 122, 432. doi:10.1002/2016JE005240

\vspace{-1em}

\bibitem[Hubbard et al.(1986)]{Neptune_rings} Hubbard, W.~B., Brahic, A., Sicardy, B., et al.\ 1986, \nat, 319, 636. doi:10.1038/319636a0

\vspace{-1em}

\bibitem[Huber et al.(2013)]{Huber:2013} Huber, D., Carter, J.~A., Barbieri, M., et al.\ 2013, Science, 342, 331. doi:10.1126/science.1242066

\vspace{-1em}

\bibitem[Hwang et al.(2018)]{Hwang:2018} Hwang, K.-H., Udalski, A., Bond, I.~A., et al.\ 2018, \aj, 155, 259. doi:10.3847/1538-3881/aac2cb

\vspace{-1em}

\bibitem[Hyodo et al.(2015)]{Hyodo:2015} Hyodo, R., Ohtsuki, K., \& Takeda, T.\ 2015, \apj, 799, 40. doi:10.1088/0004-637X/799/1/40

\vspace{-1em}

\bibitem[Hyodo \& Ohtsuki(2015)]{moon_collision} Hyodo, R. \& Ohtsuki, K.\ 2015, Nature Geoscience, 8, 686. doi:10.1038/ngeo2508

\vspace{-1em}

\bibitem[Isella et al.(2019)]{Isella:2019} Isella, A., Benisty, M., Teague, R., et al.\ 2019, \apjl, 879, L25. doi:10.3847/2041-8213/ab2a12

\vspace{-1em}

\bibitem[J{\"a}ger \& Szab{\'o}(2021)]{Jager:2021} J{\"a}ger, Z. \& Szab{\'o}, G.~M.\ 2021, \mnras, 508, 5524. doi:10.1093/mnras/stab2955

\vspace{-1em}

\bibitem[Johnson \& Huggins(2006)]{Johnson:2006} Johnson, R.~E. \& Huggins, P.~J.\ 2006, \pasp, 118, 1136. doi:10.1086/506183

\vspace{-1em}

\bibitem[Jones et al.(2006)]{Jones:2006} Jones, H.~R.~A., Butler, R.~P., Tinney, C.~G., et al.\ 2006, \mnras, 369, 249. doi:10.1111/j.1365-2966.2006.10298.x

\vspace{-1em}

\bibitem[Kaltenegger(2010)]{Kaltenegger:2010} Kaltenegger, L.\ 2010, \apjl, 712, L125. doi:10.1088/2041-8205/712/2/L125

\vspace{-1em}

\bibitem[Kawauchi et al.(2022)]{Kawauchi:2022} Kawauchi, K., Narita, N., Sato, B., et al.\ 2022, \pasj. doi:10.1093/pasj/psab120

\vspace{-1em}

\bibitem[Kenworthy \& Mamajek(2015)]{Kenworthy:2015} Kenworthy, M.~A. \& Mamajek, E.~E.\ 2015, \apj, 800, 126. doi:10.1088/0004-637X/800/2/126

\vspace{-1em}

\bibitem[Kenworthy et al.(2015)]{Kenworthy:2015b} Kenworthy, M.~A., Lacour, S., Kraus, A., et al.\ 2015, \mnras, 446, 411. doi:10.1093/mnras/stu2067 

\vspace{-1em}

\bibitem[Kenworthy et al.(2020)]{Kenworthy:2020} Kenworthy, M.~A., Klaassen, P.~D., Min, M., et al.\ 2020, \aap, 633, A115. doi:10.1051/0004-6361/201936141

\vspace{-1em}

\bibitem[Keppler et al.(2018)]{Keppler:2018} Keppler, M., Benisty, M., M{\"u}ller, A., et al.\ 2018, \aap, 617, A44. doi:10.1051/0004-6361/201832957

\vspace{-1em}

\bibitem[Khurana et al.(1998)]{Europa_ocean} Khurana, K.~K., Kivelson, M.~G., Stevenson, D.~J., et al.\ 1998, \nat, 395, 777. doi:10.1038/27394

\vspace{-1em}

\bibitem[Kipping(2009a)]{Kipping:2009} Kipping, D.~M.\ 2009, \mnras, 392, 181. doi:10.1111/j.1365-2966.2008.13999.x

\vspace{-1em}

\bibitem[Kipping(2009b)]{Kipping:2009b} Kipping, D.~M.\ 2009, \mnras, 396, 1797. doi:10.1111/j.1365-2966.2009.14869.x

\vspace{-1em}

\bibitem[Kipping et al.(2009)]{Kipping:2009c} Kipping, D.~M., Fossey, S.~J., \& Campanella, G.\ 2009, \mnras, 400, 398. doi:10.1111/j.1365-2966.2009.15472.x

\vspace{-1em}

\bibitem[Kipping(2010)]{Kipping:2010} Kipping, D.~M.\ 2010, \mnras, 408, 1758. doi:10.1111/j.1365-2966.2010.17242.x

\bibitem[Kipping(2011)]{LUNA} Kipping, D.~M.\ 2011, \mnras, 416, 689. doi:10.1111/j.1365-2966.2011.19086.x

\bibitem[Kipping et al.(2012)]{HEKI} Kipping, D.~M., Bakos, G. {\'A}., Buchhave, L., et al.\ 2012, \apj, 750, 115. doi:10.1088/0004-637X/750/2/115

\vspace{-1em}

\bibitem[Kipping et al.(2013a)]{HEKII} Kipping, D.~M., Hartman, J., Buchhave, L.~A., et al.\ 2013, \apj, 770, 101. doi:10.1088/0004-637X/770/2/101

\vspace{-1em}

\bibitem[Kipping et al.(2013b)]{HEKIII} Kipping, D.~M., Forgan, D., Hartman, J., et al.\ 2013, \apj, 777, 134. doi:10.1088/0004-637X/777/2/134

\vspace{-1em}

\bibitem[Kipping et al.(2014)]{HEKIV} Kipping, D.~M., Nesvorn{\'y}, D., Buchhave, L.~A., et al.\ 2014, \apj, 784, 28. doi:10.1088/0004-637X/784/1/28

\vspace{-1em}

\bibitem[Kipping et al.(2015)]{HEKV} Kipping, D.~M., Schmitt, A.~R., Huang, X., et al.\ 2015, \apj, 813, 14. doi:10.1088/0004-637X/813/1/14

\vspace{-1em}

\bibitem[Kipping(2020)]{Kipping:2020} Kipping, D.\ 2020, \apjl, 900, L44. doi:10.3847/2041-8213/abafa9

\vspace{-1em}

\bibitem[Kipping \& Teachey(2020)]{Kipping:2020b} Kipping, D. \& Teachey, A.\ 2020, Serbian Astronomical Journal, 201, 25. doi:10.2298/SAJ2001025K

\vspace{-1em}

\bibitem[Kipping(2021a)]{Kipping:2021} Kipping, D.\ 2021, \mnras, 500, 1851. doi:10.1093/mnras/staa3398

\vspace{-1em}

\bibitem[Kipping(2021b)]{transit_origami} Kipping, D.\ 2021, \mnras, 507, 4120. doi:10.1093/mnras/stab2013

\vspace{-1em}

\bibitem[Kipping et al.(2022)]{k1708} Kipping, D., Bryson, S., Burke, C., et al.\ 2022, Nature Astronomy. doi:10.1038/s41550-021-01539-1

\vspace{-1em}

\bibitem[Kipping \& Yahalomi(2023)]{Kipping_Yahalomi:2023} Kipping, D. \& Yahalomi, D.~A.\ 2023, \mnras, 518, 3482. doi:10.1093/mnras/stac3360

\vspace{-1em}

\bibitem[Kivelson et al.(2002)]{Ganymede_ocean} Kivelson, M.~G., Khurana, K.~K., \& Volwerk, M.\ 2002, \icarus, 157, 507. doi:10.1006/icar.2002.6834

\vspace{-1em}

\bibitem[Kleisioti et al.(2023)]{Kleisioti:2023} Kleisioti, E., Dirkx, D., Rovira-Navarro, M., et al.\ 2023, \aap, 675, A57. doi:10.1051/0004-6361/202346082

\vspace{-1em}

\bibitem[Kollmeier \& Raymond(2019)]{submoons} Kollmeier, J.~A. \& Raymond, S.~N.\ 2019, \mnras, 483, L80. doi:10.1093/mnrasl/sly219

\vspace{-1em}

\bibitem[Kov{\'a}cs et al.(2002)]{BLS} Kov{\'a}cs, G., Zucker, S., \& Mazeh, T.\ 2002, \aap, 391, 369. doi:10.1051/0004-6361:20020802

\vspace{-1em}

\bibitem[Kreidberg et al.(2019)]{Kreidberg:2019} Kreidberg, L., Luger, R., \& Bedell, M.\ 2019, \apjl, 877, L15. doi:10.3847/2041-8213/ab20c8

\vspace{-1em}

\bibitem[Kuiper(1944)]{Kuiper:1944} Kuiper, G.~P.\ 1944, \apj, 100, 378. doi:10.1086/144679

\vspace{-1em}

\bibitem[Lammer et al.(2014)]{Lammer:2014} Lammer, H., Schiefer, S.-C., Juvan, I., et al.\ 2014, Origins of Life and Evolution of the Biosphere, 44, 239. doi:10.1007/s11084-014-9377-2

\vspace{-1em}

\bibitem[Lammer et al.(2016)]{Lammer:2016} Lammer, H., Erkaev, N.~V., Fossati, L., et al.\ 2016, \mnras, 461, L62. doi:10.1093/mnrasl/slw095

\vspace{-1em}

\bibitem[Laskar \& Petit(2017)]{Laskar:2017} Laskar, J. \& Petit, A.~C.\ 2017, \aap, 605, A72. doi:10.1051/0004-6361/201630022

\vspace{-1em}

\bibitem[Law et al.(2015)]{EVRYSCOPE} Law, N.~M., Fors, O., Ratzloff, J., et al.\ 2015, \pasp, 127, 234. doi:10.1086/680521

\vspace{-1em}

\bibitem[Law et al.(2021)]{Law:2021} Law, N.~M., Corbett, H., Galliher, N.~W., et al.\ 2021, arXiv:2107.00664

\vspace{-1em}

\bibitem[Laskar \& Robutel(1993)]{moon_important} Laskar, J. \& Robutel, P.\ 1993, \nat, 361, 608. doi:10.1038/361608a0

\vspace{-1em}

\bibitem[Lecavelier des Etangs et al.(2017)]{CoRoT-9b} Lecavelier des Etangs, A., H{\'e}brard, G., Blandin, S., et al.\ 2017, \aap, 603, A115. doi:10.1051/0004-6361/201730554

\vspace{-1em}

\bibitem[Levison et al.(2008)]{kuiper_belt_origin} Levison, H.~F., Morbidelli, A., Van Laerhoven, C., et al.\ 2008, \icarus, 196, 258. doi:10.1016/j.icarus.2007.11.035

\vspace{-1em}

\bibitem[Lewis et al.(2008)]{Lewis:2008} Lewis, K.~M., Sackett, P.~D., \& Mardling, R.~A.\ 2008, \apjl, 685, L153. doi:10.1086/592743

\vspace{-1em}

\bibitem[Limbach et al.(2021)]{Limbach:2021} Limbach, M.~A., Vos, J.~M., Winn, J.~N., et al.\ 2021, \apjl, 918, L25. doi:10.3847/2041-8213/ac1e2d

\vspace{-1em}

\bibitem[Limbach et al.(2023)]{Limbach:2023} Limbach, M.~A., Soares-Furtado, M., Vanderburg, A., et al.\ 2023, \pasp, 135, 014401. doi:10.1088/1538-3873/acafa4

\vspace{-1em}

\bibitem[Lissauer et al.(2012)]{moon_not_important} Lissauer, J.~J., Barnes, J.~W., \& Chambers, J.~E.\ 2012, \icarus, 217, 77. doi:10.1016/j.icarus.2011.10.013

\vspace{-1em}

\bibitem[Lorenz et al.(2008)]{Titan_ocean} Lorenz, R.~D., Stiles, B.~W., Kirk, R.~L., et al.\ 2008, Science, 319, 1649. doi:10.1126/science.1151639


\bibitem[Lucey et al.(1995)]{Lucey:1995} Lucey, P.~G., Taylor, G.~J., \& Malaret, E.\ 1995, Science, 268, 1150. doi:10.1126/science.268.5214.1150

\vspace{-1em}

\bibitem[Luger et al.(2017)]{Luger:2017} Luger, R., Sestovic, M., Kruse, E., et al.\ 2017, Nature Astronomy, 1, 0129. doi:10.1038/s41550-017-0129

\vspace{-1em}

\bibitem[Lunine \& Stevenson(1982)]{Lunine:1982} Lunine, J.~I. \& Stevenson, D.~J.\ 1982, \icarus, 52, 14. doi:10.1016/0019-1035(82)90166-X

\vspace{-1em}

\bibitem[Makarov \& Efroimsky(2023)]{Makarov:2023} Makarov, V.~V. \& Efroimsky, M.\ 2023, \aap, 672, A78. doi:10.1051/0004-6361/202245533

\vspace{-1em}

\bibitem[Malamud et al.(2020)]{Malamud:2020} Malamud, U., Perets, H.~B., Sch{\"a}fer, C., et al.\ 2020, \mnras, 492, 5089. doi:10.1093/mnras/staa211

\vspace{-1em}

\bibitem[Mandel \& Agol(2002)]{MandelAgol:2002} Mandel, K. \& Agol, E.\ 2002, \apjl, 580, L171. doi:10.1086/345520

\vspace{-1em}

\bibitem[Marois et al.(2008)]{Marois:2008} Marois, C., Macintosh, B., Barman, T., et al.\ 2008, Science, 322, 1348. doi:10.1126/science.1166585

\vspace{-1em}

\bibitem[Martinez et al.(2019)]{Martinez:2019} Martinez, M.~A.~S., Stone, N.~C., \& Metzger, B.~D.\ 2019, \mnras, 489, 5119. doi:10.1093/mnras/stz2464

\vspace{-1em}

\bibitem[Martin et al.(2019)]{Martin:2019} Martin, D.~V., Fabrycky, D.~C., \& Montet, B.~T.\ 2019, \apjl, 875, L25. doi:10.3847/2041-8213/ab0aea

\vspace{-1em}

\bibitem[Martins et al.(2015)]{Martins:2015} Martins, J.~H.~C., Santos, N.~C., Figueira, P., et al.\ 2015, \aap, 576, A134. doi:10.1051/0004-6361/201425298

\vspace{-1em}

\bibitem[Mayor \& Queloz(1995)]{Mayor:1995} Mayor, M. \& Queloz, D.\ 1995, \nat, 378, 355. doi:10.1038/378355a0

\vspace{-1em}

\bibitem[Mentel et al.(2018)]{Mentel:2018} Mentel, R.~T., Kenworthy, M.~A., Cameron, D.~A., et al.\ 2018, \aap, 619, A157. doi:10.1051/0004-6361/201834004 

\vspace{-1em}

\bibitem[Metzger et al.(2022)]{Metzger:2022} Metzger, P.~T., Grundy, W.~M., Sykes, M.~V., et al.\ 2022, \icarus, 374, 114768. doi:10.1016/j.icarus.2021.114768

\vspace{-1em}

\bibitem[Millholland(2019)]{Millholland:2019} Millholland, S.\ 2019, \apj, 886, 72. doi:10.3847/1538-4357/ab4c3f

\vspace{-1em}

\bibitem[Miyazaki et al.(2018)]{Miyazaki:2018} Miyazaki, S., Sumi, T., Bennett, D.~P., et al.\ 2018, \aj, 156, 136. doi:10.3847/1538-3881/aad5ee

\vspace{-1em}

\bibitem[Moraes \& Vieira Neto(2020)]{Moraes:2020} Moraes, R.~A. \& Vieira Neto, E.\ 2020, \mnras, 495, 3763. doi:10.1093/mnras/staa1441

\vspace{-1em}

\bibitem[Moraes et al.(2022)]{Moraes:2022} Moraes, R.~A., Borderes-Motta, G., Winter, O.~C., et al.\ 2022, \mnras, 510, 2583. doi:10.1093/mnras/stab3576

\vspace{-1em}

\bibitem[Morbidelli \& Crida(2007)]{Morbidelli:2007} Morbidelli, A. \& Crida, A.\ 2007, \icarus, 191, 158. doi:10.1016/j.icarus.2007.04.001

\vspace{-1em}

\bibitem[Morgado et al.(2023)]{Quaoar_rings} Morgado, B.~E., Sicardy, B., Braga-Ribas, F., et al.\ 2023, \nat, 614, 239. doi:10.1038/s41586-022-05629-6

\vspace{-1em}

\bibitem[Mosqueira \& Estrada(2003)]{Mosqueira:2003} Mosqueira, I. \& Estrada, P.~R.\ 2003, \icarus, 163, 232. doi:10.1016/S0019-1035(03)00077-0

\vspace{-1em}

\bibitem[Nakajima et al.(2022)]{Nakajima:2022} Nakajima, M., Genda, H., Asphaug, E., et al.\ 2022, Nature Communications, 13, 568. doi:10.1038/s41467-022-28063-8

\vspace{-1em}

\bibitem[Namouni(2010)]{Namouni:2010} Namouni, F.\ 2010, \apjl, 719, L145

\vspace{-1em}

\bibitem[Narang et al.(2023a)]{Narang:2023a} Narang, M., Oza, A.~V., Hakim, K., et al.\ 2023, \aj, 165, 1. doi:10.3847/1538-3881/ac9eb8

\vspace{-1em}

\bibitem[Narang et al.(2023b)]{Narang:2023b} Narang, M., Oza, A.~V., Hakim, K., et al.\ 2023, \mnras, 522, 1662. doi:10.1093/mnras/stad1027

\vspace{-1em}

\bibitem[Nesvorn{\'y} et al.(2007)]{Nesvorny:2007} Nesvorn{\'y}, D., Vokrouhlick{\'y}, D., \& Morbidelli, A.\ 2007, \aj, 133, 1962

\vspace{-1em}

\bibitem[Noyola et al.(2014)]{Noyola:2014} Noyola, J.~P., Satyal, S., \& Musielak, Z.~E.\ 2014, \apj, 791, 25. doi:10.1088/0004-637X/791/1/25

\vspace{-1em}

\bibitem[Noyola et al.(2016)]{Noyola:2016} Noyola, J.~P., Satyal, S., \& Musielak, Z.~E.\ 2016, \apj, 821, 97. doi:10.3847/0004-637X/821/2/97

\vspace{-1em}

\bibitem[O'Brien et al.(2014)]{Obrien:2014} O'Brien, D.~P., Walsh, K.~J., Morbidelli, A., et al.\ 2014, \icarus, 239, 74. doi:10.1016/j.icarus.2014.05.009

\vspace{-1em}

\bibitem[O'Donoghue et al.(2019)]{Saturn_ring_loss} O'Donoghue, J., Moore, L., Connerney, J., et al.\ 2019, \icarus, 322, 251. doi:10.1016/j.icarus.2018.10.027

\vspace{-1em}

\bibitem[Ogihara \& Ida(2012)]{Ogihara:2012} Ogihara, M. \& Ida, S.\ 2012, \apj, 753, 60. doi:10.1088/0004-637X/753/1/60

\vspace{-1em}

\bibitem[Ohno \& Fortney(2022)]{Ohno:2022} Ohno, K. \& Fortney, J.~J.\ 2022, \apj, 930, 50. doi:10.3847/1538-4357/ac6029

\vspace{-1em}

\bibitem[Ohno et al.(2022)]{Ohno:2022b} Ohno, K., Thao, P.~C., Mann, A.~W., et al.\ 2022, \apjl, 940, L30. doi:10.3847/2041-8213/ac9f3f

\vspace{-1em}

\bibitem[Ohta et al.(2009)]{Ohta:2009} Ohta, Y., Taruya, A., \& Suto, Y.\ 2009, \apj, 690, 1. doi:10.1088/0004-637X/690/1/1

\vspace{-1em}

\bibitem[Ortiz et al.(2015)]{Chiron_rings} Ortiz, J.~L., Duffard, R., Pinilla-Alonso, N., et al.\ 2015, \aap, 576, A18. doi:10.1051/0004-6361/201424461

\vspace{-1em}

\bibitem[Ortiz et al.(2017)]{Haumea_rings} Ortiz, J.~L., Santos-Sanz, P., Sicardy, B., et al.\ 2017, \nat, 550, 219. doi:10.1038/nature24051

\vspace{-1em}

\bibitem[Osborn et al.(2016)]{Osborn:2016} Osborn, H.~P., Armstrong, D.~J., Brown, D.~J.~A., et al.\ 2016, \mnras, 457, 2273. doi:10.1093/mnras/stw137

\vspace{-1em}

\bibitem[Osborn et al.(2019)]{Osborn:2019} Osborn, H.~P., Ansdell, M., Ioannou, Y., et al.\ 2019, arXiv e-prints, arXiv:1902.08544

\vspace{-1em}

\bibitem[Owen(1965)]{Saturn_ring_ice} Owen, T.\ 1965, Science, 149, 974. doi:10.1126/science.149.3687.974

\vspace{-1em}

\bibitem[Oza et al.(2019)]{Oza:2019} Oza, A.~V., Johnson, R.~E., Lellouch, E., et al.\ 2019, \apj, 885, 168. doi:10.3847/1538-4357/ab40cc

\vspace{-1em}

\bibitem[Pasqua \& Assaf(2014)]{Pasqua:2014} Pasqua, A. \& Assaf, K.~A.\ 2014, Advances in Astronomy, 2014, 450864. doi:10.1155/2014/450864

\vspace{-1em}

\bibitem[Perrocheau et al.(2022)]{unistellar_k167} Perrocheau, A., Esposito, T.~M., Dalba, P.~A., et al.\ 2022, \apjl, 940, L39. doi:10.3847/2041-8213/aca073

\vspace{-1em}

\bibitem[Peters-Limbach \& Turner(2013)]{Limbach:2013} Peters-Limbach, M.~A. \& Turner, E.~L.\ 2013, \apj, 769, 98. doi:10.1088/0004-637X/769/2/98

\vspace{-1em}

\bibitem[Petit et al.(2018)]{Petit:2018} Petit, A.~C., Laskar, J., \& Bou{\'e}, G.\ 2018, \aap, 617, A93. doi:10.1051/0004-6361/201833088

\vspace{-1em}

\bibitem[Piro \& Vissapragada(2020)]{Piro:2020} Piro, A.~L. \& Vissapragada, S.\ 2020, \aj, 159, 131. doi:10.3847/1538-3881/ab7192

\vspace{-1em}

\bibitem[Pollack et al.(1979)]{Pollack:1979} Pollack, J.~B., Burns, J.~A., \& Tauber, M.~E.\ 1979, \icarus, 37, 587. doi:10.1016/0019-1035(79)90016-2

\vspace{-1em}

\bibitem[Porco et al.(2006)]{Porco:2006} Porco, C.~C., Helfenstein, P., Thomas, P.~C., et al.\ 2006, Science, 311, 1393. doi:10.1126/science.1123013

\vspace{-1em}

\bibitem[Postberg et al.(2009)]{Enceladus_ocean} Postberg, F., Kempf, S., Schmidt, J., et al.\ 2009, \nat, 459, 1098. doi:10.1038/nature08046

\vspace{-1em}

\bibitem[Quarles et al.(2020)]{Quarles:2020} Quarles, B., Li, G., \& Rosario-Franco, M.\ 2020, \apjl, 902, L20. doi:10.3847/2041-8213/abba36

\vspace{-1em}

\bibitem[Rabago \& Steffen(2019)]{Rabago:2019} Rabago, I. \& Steffen, J.~H.\ 2019, \mnras, 489, 2323. doi:10.1093/mnras/sty2552

\vspace{-1em}

\bibitem[Ragozzine \& Wolf(2009)]{Ragozzine:2009} Ragozzine, D. \& Wolf, A.~S.\ 2009, \apj, 698, 1778. doi:10.1088/0004-637X/698/2/1778

\vspace{-1em}

\bibitem[Rein \& Liu(2012)]{Rein:2012} Rein, H. \& Liu, S.-F.\ 2012, \aap, 537, A128. doi:10.1051/0004-6361/201118085

\vspace{-1em}

\bibitem[Reynolds et al.(1983)]{Reynolds:1983} Reynolds, R.~T., Squyres, S.~W., Colburn, D.~S., et al.\ 1983, \icarus, 56, 246. doi:10.1016/0019-1035(83)90037-4

\vspace{-1em}

\bibitem[Riaud et al.(2006)]{Riaud:2006} Riaud, P., Mawet, D., Absil, O., et al.\ 2006, \aap, 458, 317. doi:10.1051/0004-6361:20065232

\vspace{-1em}

\bibitem[Rieder \& Kenworthy(2016)]{Rieder:2016} Rieder, S. \& Kenworthy, M.~A.\ 2016, \aap, 596, A9. doi:10.1051/0004-6361/201629567

\vspace{-1em}

\bibitem[Rodenbeck et al.(2018)]{Rodenbeck:2018} Rodenbeck, K., Heller, R., Hippke, M., et al.\ 2018, \aap, 617, A49. doi:10.1051/0004-6361/201833085

\vspace{-1em}

\bibitem[Rodenbeck et al.(2020)]{Rodenbeck:2020} Rodenbeck, K., Heller, R., \& Gizon, L.\ 2020, \aap, 638, A43. doi:10.1051/0004-6361/202037550

\vspace{-1em}

\bibitem[Roth et al.(2014)]{Europa_geyser1} Roth, L., Saur, J., Retherford, K.~D., et al.\ 2014, Science, 343, 171. doi:10.1126/science.1247051

\vspace{-1em}

\bibitem[Rosario-Franco et al.(2020)]{RosarioFranco:2020} Rosario-Franco, M., Quarles, B., Musielak, Z.~E., et al.\ 2020, \aj, 159, 260. doi:10.3847/1538-3881/ab89a7

\vspace{-1em}

\bibitem[Roth et al.(2014)]{Roth:2014} Roth, L., Saur, J., Retherford, K.~D., et al.\ 2014, Science, 343, 171. doi:10.1126/science.1247051

\vspace{-1em}

\bibitem[Ruffio et al.(2023)]{Ruffio:2023} Ruffio, J.-B., Horstman, K., Mawet, D., et al.\ 2023, \aj, 165, 113. doi:10.3847/1538-3881/acb34a

\vspace{-1em}

\bibitem[Saha \& Sengupta(2022)]{Saha:2022} Saha, S. \& Sengupta, S.\ 2022, \apj, 936, 2. doi:10.3847/1538-4357/ac85a9

\vspace{-1em}

\bibitem[Sandford \& Kipping(2017)]{Sandford:2017} Sandford, E., \& Kipping, D.\ 2017, \aj, 154, 228

\vspace{-1em}

\bibitem[Sandford et al.(2019)]{Sandford:2019} Sandford, E., Espinoza, N., Brahm, R., et al.\ 2019, arXiv e-prints, arXiv:1908.08548

\vspace{-1em}

\bibitem[Santos et al.(2015)]{Santos:2015} Santos, N.~C., Martins, J.~H.~C., Bou{\'e}, G., et al.\ 2015, \aap, 583, A50. doi:10.1051/0004-6361/201526673

\vspace{-1em}

\bibitem[Sartoretti \& Schneider(1999)]{Sartoretti:1999} Sartoretti, P. \& Schneider, J.\ 1999, \aaps, 134, 553. doi:10.1051/aas:1999148

\vspace{-1em}

\bibitem[Sasaki et al.(2012)]{Sasaki:2012} Sasaki, T., Barnes, J.~W., \& O'Brien, D.~P.\ 2012, \apj, 754, 51. doi:10.1088/0004-637X/754/1/51

\vspace{-1em}

\bibitem[Sasaki \& Hosono(2018)]{Sasaki:2018} Sasaki, T. \& Hosono, N.\ 2018, \apj, 856, 175. doi:10.3847/1538-4357/aab369

\vspace{-1em}

\bibitem[Scharf(2006)]{Scharf:2006} Scharf, C.~A.\ 2006, \apj, 648, 1196. doi:10.1086/505256

\vspace{-1em}

\bibitem[Shabram \& Boley(2013)]{Shabram:2013} Shabram, M. \& Boley, A.~C.\ 2013, \apj, 767, 63. doi:10.1088/0004-637X/767/1/63

\vspace{-1em}

\bibitem[Shallue \& Vanderburg(2018)]{Shallue:2018} Shallue, C.~J., \& Vanderburg, A.\ 2018, \aj, 155, 94

\vspace{-1em}

\bibitem[Showalter et al.(1991)]{Enceladus_Ering} Showalter, M.~R., Cuzzi, J.~N., \& Larson, S.~M.\ 1991, \icarus, 94, 451. doi:10.1016/0019-1035(91)90241-K

\vspace{-1em}

\bibitem[Simon et al.(2007)]{Simon:2007} Simon, A., Szatm{\'a}ry, K., \& Szab{\'o}, G.~M.\ 2007, \aap, 470, 727. doi:10.1051/0004-6361:20066560

\vspace{-1em}

\bibitem[Simon et al.(2010)]{Simon:2010} Simon, A.~E., Szab{\'o}, G.~M., Szatm{\'a}ry, K., et al.\ 2010, \mnras, 406, 2038. doi:10.1111/j.1365-2966.2010.16818.x

\vspace{-1em}

\bibitem[Smith et al.(1979)]{Io_volcanoes} Smith, B.~A., Soderblom, L.~A., Johnson, T.~V., et al.\ 1979, Science, 204, 951. doi:10.1126/science.204.4396.951

\vspace{-1em}

\bibitem[Smith et al.(1989)]{Neptune_ring2} Smith, B.~A., Soderblom, L.~A., Banfield, D., et al.\ 1989, Science, 246, 1422. doi:10.1126/science.246.4936.1422

\vspace{-1em}

\bibitem[Sparks et al.(2016)]{Europa_geyser2} Sparks, W.~B., Hand, K.~P., McGrath, M.~A., et al.\ 2016, \apj, 829, 121. doi:10.3847/0004-637X/829/2/121

\vspace{-1em}

\bibitem[Speedie \& Zanazzi(2020)]{Speedie:2020} Speedie, J. \& Zanazzi, J.~J.\ 2020, \mnras, 497, 1870. doi:10.1093/mnras/staa2068

\vspace{-1em}

\bibitem[\protect\citeauthoryear{Spergel et al.}{2015}]{Spergel:2015} Spergel D., Gehrels N., Baltay C., Bennett D., Breckinridge J., Donahue M., Dressler A., et al., 2015, arXiv, arXiv:1503.03757

\vspace{-1em}

\bibitem[Spalding et al.(2016)]{Spalding:2016} Spalding, C., Batygin, K., \& Adams, F.~C.\ 2016, \apj, 817, 18

\vspace{-1em}

\bibitem[Spiegel et al.(2011)]{Spiegel:2011} Spiegel, D.~S., Burrows, A., \& Milsom, J.~A.\ 2011, \apj, 727, 57. doi:10.1088/0004-637X/727/1/57

\vspace{-1em}

\bibitem[Squyres et al.(1983)]{Squyres:1983} Squyres, S.~W., Reynolds, R.~T., \& Cassen, P.~M.\ 1983, \nat, 301, 225. doi:10.1038/301225a0

\vspace{-1em}

\bibitem[Stern et al.(2006)]{Stern:2006} Stern, S.~A., Weaver, H.~A., Steffl, A.~J., et al.\ 2006, \nat, 439, 946. doi:10.1038/nature04548

\vspace{-1em}

\bibitem[Stofan et al.(2007)]{Titan_lakes} Stofan, E.~R., Elachi, C., Lunine, J.~I., et al.\ 2007, \nat, 445, 61. doi:10.1038/nature05438

\vspace{-1em}

\bibitem[Sucerquia et al.(2017)]{Sucerquia:2017} Sucerquia, M., Alvarado-Montes, J.~A., Ram{\'\i}rez, V., et al.\ 2017, \mnras, 472, L120. doi:10.1093/mnrasl/slx151

\vspace{-1em}

\bibitem[Sucerquia et al.(2019)]{Sucerquia:2019} Sucerquia, M., Alvarado-Montes, J.~A., Zuluaga, J.~I., et al.\ 2019, \mnras, 489, 2313. doi:10.1093/mnras/stz2110

\vspace{-1em}

\bibitem[Sucerquia et al.(2020)]{Sucerquia:2020} Sucerquia, M., Ram{\'\i}rez, V., Alvarado-Montes, J.~A., et al.\ 2020, \mnras, 492, 3499. doi:10.1093/mnras/stz3548

\vspace{-1em}

\bibitem[Sucerquia et al.(2022)]{cronomoons} Sucerquia, M., Alvarado-Montes, J.~A., Bayo, A., et al.\ 2022, \mnras, 512, 1032. doi:10.1093/mnras/stab3531

\vspace{-1em}

\bibitem[Sutton(2019)]{Sutton:2019} Sutton, P.~J.\ 2019, \mnras, 486, 1681. doi:10.1093/mnras/stz563

\vspace{-1em}

\bibitem[Sutton et al.(2022)]{Sutton:2022} Sutton, P.~J., Albery, B., \& Muff, J.\ 2022, Frontiers in Astronomy and Space Sciences, 9, 819933. doi:10.3389/fspas.2022.819933

\vspace{-1em}

\bibitem[Szab{\'o} et al.(2006)]{Szabo:2006} Szab{\'o}, G.~M., Szatm{\'a}ry, K., Div{\'e}ki, Z., et al.\ 2006, \aap, 450, 395. doi:10.1051/0004-6361:20054555

\vspace{-1em}

\bibitem[Szul{\'a}gyi et al.(2016)]{Szulagyi:2016} Szul{\'a}gyi, J., Masset, F., Lega, E., et al.\ 2016, \mnras, 460, 2853. doi:10.1093/mnras/stw1160

\vspace{-1em}

\bibitem[Szul{\'a}gyi(2017)]{Szulagyi:2017} Szul{\'a}gyi, J.\ 2017, \apj, 842, 103. doi:10.3847/1538-4357/aa7515

\vspace{-1em}

\bibitem[Szul{\'a}gyi et al.(2017)]{Szulagyi:2017b} Szul{\'a}gyi, J., Mayer, L., \& Quinn, T.\ 2017, \mnras, 464, 3158. doi:10.1093/mnras/stw2617

\vspace{-1em}

\bibitem[Szul{\'a}gyi et al.(2018)]{Szulagyi:2018} Szul{\'a}gyi, J., Cilibrasi, M., \& Mayer, L.\ 2018, \apjl, 868, L13. doi:10.3847/2041-8213/aaeed6

\vspace{-1em}

\bibitem[Szul{\'a}gyi et al.(2022)]{Szulagyi:2022} Szul{\'a}gyi, J., Binkert, F., \& Surville, C.\ 2022, \apj, 924, 1. doi:10.3847/1538-4357/ac32d1

\vspace{-1em}

\bibitem[Tamayo et al.(2020)]{Tamayo:2020} Tamayo, D., Cranmer, M., Hadden, S., et al.\ 2020, Proceedings of the National Academy of Science, 117, 18194. doi:10.1073/pnas.2001258117

\vspace{-1em}

\bibitem[Teachey et al.(2018)]{HEKVI} Teachey, A., Kipping, D.~M., \& Schmitt, A.~R.\ 2018, \aj, 155, 36

\vspace{-1em}

\bibitem[Teachey \& Kipping(2018)]{k1625_a} Teachey, A., \& Kipping, D.~M.\ 2018, \textit{Science Advances}, 4, eaav1784

\vspace{-1em}

\bibitem[Teachey et al.(2019)]{k1625_b} Teachey, A., Kipping, D., Burke, C.~J., et al.\ 2019, arXiv e-prints, arXiv:1904.11896

\vspace{-1em}

\bibitem[Teachey(2021)]{multimoons} Teachey, A.\ 2021, \mnras, 506, 2104. doi:10.1093/mnras/stab1840

\vspace{-1em}

\bibitem[Teachey \& Kipping(2021)]{CNNs} Teachey, A. \& Kipping, D.\ 2021, \mnras. doi:10.1093/mnras/stab2694

\vspace{-1em}

\bibitem[Thao et al.(2023)]{Thao:2023} Thao, P.~C., Mann, A.~W., Gao, P., et al.\ 2023, \aj, 165, 23. doi:10.3847/1538-3881/aca07a

\vspace{-1em}

\bibitem[Timmermann et al.(2020)]{Timmermann:2020} Timmermann, A., Heller, R., Reiners, A., et al.\ 2020, \aap, 635, A59. doi:10.1051/0004-6361/201937325

\vspace{-1em}

\bibitem[Tjoa et al.(2020)]{Tjoa:2020} Tjoa, J.~N.~K.~Y., Mueller, M., \& van der Tak, F.~F.~S.\ 2020, \aap, 636, A50. doi:10.1051/0004-6361/201937035

\vspace{-1em}

\bibitem[Tokadjian \& Piro(2020)]{Tokadjian:2020} Tokadjian, A. \& Piro, A.~L.\ 2020, \aj, 160, 194. doi:10.3847/1538-3881/abb29e

\vspace{-1em}

\bibitem[Tokadjian \& Piro(2023)]{Tokadjian:2023} Tokadjian, A. \& Piro, A.~L.\ 2023, \aj, 165, 173. doi:10.3847/1538-3881/acc254

\vspace{-1em}

\bibitem[Trani et al.(2020)]{Trani:2020} Trani, A.~A., Hamers, A.~S., Geller, A., et al.\ 2020, \mnras, 499, 4195. doi:10.1093/mnras/staa3098

\vspace{-1em}

\bibitem[Vanderburg et al.(2018)]{Vanderburg:2018} Vanderburg, A., Rappaport, S.~A., \& Mayo, A.~W.\ 2018, \aj, 156, 184. doi:10.3847/1538-3881/aae0fc

\vspace{-1em}

\bibitem[\protect\citeauthoryear{Vanderburg \& Rodriguez}{2021}]{Vanderburg:2021} Vanderburg A., Rodriguez J.~E., 2021, arXiv, arXiv:2110.14650

\vspace{-1em}

\bibitem[Vance et al.(2018)]{Vance:2018} Vance, S.~D., Panning, M.~P., St{\"a}hler, S., et al.\ 2018, Journal of Geophysical Research (Planets), 123, 180. doi:10.1002/2017JE005341

\vspace{-1em}

\bibitem[Verbiscer et al.(2009)]{Phoebe_ring} Verbiscer, A.~J., Skrutskie, M.~F., \& Hamilton, D.~P.\ 2009, \nat, 461, 1098. doi:10.1038/nature08515

\vspace{-1em}

\bibitem[Weaver et al.(1995)]{Shoemaker-Levy9} Weaver, H.~A., A'Hearn, M.~F., Arpigny, C., et al.\ 1995, Science, 267, 1282. doi:10.1126/science.7871424

\vspace{-1em}

\bibitem[Williams \& Murray(2011)]{Janus_Epimetheus_ring} Williams, G.~A. \& Murray, C.~D.\ 2011, \icarus, 212, 275. doi:10.1016/j.icarus.2010.11.038

\vspace{-1em}

\bibitem[Wilson et al.(2002)]{Wilson:2002} Wilson, J.~K., Mendillo, M., Baumgardner, J., et al.\ 2002, \icarus, 157, 476. doi:10.1006/icar.2002.6821

\vspace{-1em}

\bibitem[Winn et al.(2005)]{Winn:2005} Winn, J.~N., Noyes, R.~W., Holman, M.~J., et al.\ 2005, \apj, 631, 1215. doi:10.1086/432571

\vspace{-1em}

\bibitem[Wisdom et al.(2022)]{Saturn_chrysalis} Wisdom, J., Dbouk, R., Militzer, B., et al.\ 2022, Science, 377, 1285. doi:10.1126/science.abn1234

\vspace{-1em}

\bibitem[Wolszczan \& Frail(1992)]{Wolszczan:1992} Wolszczan, A. \& Frail, D.~A.\ 1992, \nat, 355, 145. doi:10.1038/355145a0

\vspace{-1em}

\bibitem[Zhuang et al.(2012)]{Zhuang:2012} Zhuang, Q., Gao, X., \& Yu, Q.\ 2012, \apj, 758, 111. doi:10.1088/0004-637X/758/2/111

\vspace{-1em}

\bibitem[Zollinger et al.(2017)]{Zollinger:2017} Zollinger, R.~R., Armstrong, J.~C., \& Heller, R.\ 2017, \mnras, 472, 8. doi:10.1093/mnras/stx1861

\vspace{-1em}

\bibitem[Zuluaga et al.(2015)]{Zuluaga:2015} Zuluaga, J.~I., Kipping, D.~M., Sucerquia, M., et al.\ 2015, \apjl, 803, L14. doi:10.1088/2041-8205/803/1/L14

\vspace{-1em}

\bibitem[Zuluaga et al.(2022)]{Zuluaga:2022} Zuluaga, J.~I., Sucerquia, M., \& Alvarado-Montes, J.~A.\ 2022, Astronomy and Computing, 40, 100623. doi:10.1016/j.ascom.2022.100623



%\end{comment}
\end{thebibliography}
